\documentclass[dvipdfmx,10pt,a4paper]{article}  % local use(fast output)

\usepackage{amsmath,amsthm,amssymb,mathrsfs,bm,bbm,enumerate,physics}
\usepackage{tikz}
\usetikzlibrary{intersections,calc,arrows.meta}
\usepackage{float,placeins,adjustbox}
\usepackage[hyphens]{url}
\usepackage{hyperref}
\usepackage{cleveref}

\usepackage{comment}
\usepackage{caption}
\newcommand\myshade{75}
\colorlet{mylinkcolor}{red}
\colorlet{mycitecolor}{green}
\colorlet{myurlcolor}{blue}
\hypersetup{colorlinks=true, anchorcolor=cyan, 
linkcolor=mylinkcolor!\myshade!black, 
citecolor=mycitecolor!\myshade!black,
urlcolor=myurlcolor!\myshade!black } 

\newcommand{\beq}{\begin{equation}}
\newcommand{\eeq}{\end{equation}}
\newcommand{\nn}{\nonumber}
\newcommand{\ra}{\rightarrow}
\newcommand{\lra}{\leftrightarrow}

\newcommand{\eps}{\epsilon}

\newcommand{\Op}[3]{ \hat{#1}_{#2}^{({#3})}}

\newcommand{\Exp}[1]{\exp\left({#1}\right)}
\newcommand{\Tm}[1]{{\mathscr T}_{\bm{#1} } }

\theoremstyle{definition}
\newtheorem*{rmk*}{$\ddagger$  Remark}

\theoremstyle{remark}
\newtheorem*{exm*}{$\spadesuit$ Example}

\makeatletter
\long\def\@makefntext#1{\parindent 1em\noindent
\@hangfrom{\hbox to 1.8em{\hss$^{\@thefnmark}$}}#1}
\makeatother
\begin{document}
\topmargin 0pt
\oddsidemargin 0mm
\renewcommand{\thefootnote}{\fnsymbol{footnote}}

%\begin{flushright}
%manuscript 0.5 \\
%CZ-alg-003
%\end{flushright}
\vspace*{0.5cm}

\begin{center}
{\Large Gauge Invariant and Generic Formulation of Magnetic Translations and 
$\mathfrak{so}(3,1)$ Curtright-Zachos Generators}
\vspace{1.5cm}

{\large Haru-Tada Sato${\,}^{a,b}$
\footnote{\,\,Corresponding author.  E-mail address: satoh@isfactory.co.jp \\
${}^{a}$\,\,This research was completed during the author's tenure as a visiting researcher there, ending on October 31, 2024.}
}\\

{${}^{a}$\em Department of Physics, Graduate School of Science, \\Osaka Metropolitan University \\
Nakamozu Campus, Sakai, Osaka 599-8531, Japan}\\

{${}^{b}$\em Department of Data Science, i's Factory Corporation, Ltd.\\
     Kanda-nishiki-cho 2-7-6, Tokyo 101-0054, Japan}\\
%
%\vspace{0.5cm}
%This article is registered under preprint number: /hep-th/*******'.
\end{center}

\vspace{0.1cm}

\abstract{
We propose a gauge-invariant formulation of magnetic translation (GMT) operators, eliminating the gauge dependence that conventional definitions suffer from due to specific gauge choices. We extend this framework by incorporating $\mathfrak{so}(3,1)$ conformal symmetry, demonstrating how GMT operators can be naturally embedded within this algebraic structure. 
We then show the fundamental role of $\mathfrak{sim}(2)$ duality between pseudo-dilatation and pseudo-angular momentum operators in constructing GMT operators. 
A key result of this approach is the derivation of the Curtright-Zachos (CZ) generators from gauge-invariant $\mathfrak{so}(3,1)$ operators, providing a novel perspective on their algebraic properties such as the relationship between FFZ internal symmetry and $\mathfrak{sim}(2)$ duality.
}

\vspace{0.5cm}

\begin{description}
\item[Keywords:] Landau problem, gauge invariance, magnetic translation, pseudo angular momentum, Virasoro algebra 
\item[MSC:] 17B61, 17B69, 81R50, 81R60, 81V70
%\item[arXiv]: ****.****
\end{description}

%\footnote{}
%
%
%%%%%%%%%%%%%%%%%%%%  INDEX  %%%%%%%%%%%%%%%%%%%%%%%%%%%%%%%%%%%%%%%%
%\newpage
%\tableofcontents
%\setcounter{page}{1}
%\renewcommand{\thepage}{\roman{page}}

%%%%%%%%%%%%%%%%%%%%%%%%%%%%%%%%%%%%%%%%%%%%%%%%%%%%%%%%%%%%%%%%%%%%%
%                           1.     Introduction
%%%%%%%%%%%%%%%%%%%%%%%%%%%%%%%%%%%%%%%%%%%%%%%%%%%%%%%%%%%%%%%%%%%%%
\newpage
\setcounter{page}{1}
\setcounter{footnote}{0}
\setcounter{equation}{0}
\setcounter{secnumdepth}{4}
\renewcommand{\thepage}{\arabic{page}}
\renewcommand{\thefootnote}{\arabic{footnote})}
\renewcommand{\theequation}{\thesection.\arabic{equation}}

\section{Introduction}
\subsection{Background and significance}
\indent

In condensed matter physics, two-dimensional electron systems in uniform magnetic fields (Landau problem) have been extensively studied over many years as stages for diverse physical phenomena including quantum Hall effects~\cite{QHE,QHE2}, magnetoresistance effects, and superconductivity. Under ideal conditions, where effects of electron interactions and impurity scattering can be neglected, this system can be solved analytically, providing an excellent model for understanding fundamental concepts of quantum mechanics. Furthermore, the Landau problem is deeply connected to various fields of modern mathematics such as noncommutative geometry~\cite{NCG,NCFT}, quantum groups~\cite{QG}, and conformal field theory~\cite{cft1,cft2,cft3}, and its mathematical structure has influenced a wide range of theoretical physics.

The most characteristic feature of the Landau problem~\cite{LP} is the quantization of electron energy levels into discrete Landau levels. This quantization arises from the interference between the Lorentz force acting on electrons moving in a magnetic field and the wave nature of electrons. Electron motion can be decomposed into cyclotron motion  and the translational motion of the center of this circular motion (guiding center). The cyclotron motion is described quantum mechanically as a harmonic oscillator, with its energy levels corresponding to Landau levels. The motion of the guiding center, on the other hand, is described by magnetic translation operators.

Magnetic translation (MT) operators~\cite{MT} represent the phase changes experienced by electrons during translation in a magnetic field and, constitute conserved quantities of the system since they commute with the Hamiltonian. These operators play an essential role in understanding transitions between different Landau levels and the degeneracy of electron states. 
However, conventional definitions of MT operators suffer from a fundamental problem of gauge dependence, and research has been conducted to maintain gauge invariance by examining special conditions in topological materials~\cite{2dTIs}. From a symmetry structure perspective, this problem can be reframed as how to eliminate the inconvenience that arises when the algebraic relations of MT operators (commutation and composition rules) change with gauge transformations of the vector potential describing the magnetic field. When symmetry verification based on algebraic calculations is influenced by gauge choice or requires special conditions, this introduces ambiguity in physical interpretations and raises concerns about mathematical consistency through the introduction of complex structures or limitations.

The issue of gauge dependence also exists for angular momentum operators. The orbital angular momentum (OAM) of electrons is a fundamental physical quantity in atomic and nuclear physics, playing an important role in understanding the structure, spectra, and reactions of atoms and nuclei. However, since conventional angular momentum operators are not invariant under gauge transformations, they are difficult to interpret appropriately as physical quantities in electron systems in magnetic fields. To resolve this problem, gauge-invariant orbital angular momentum operators (pseudo OAM) have been introduced, and research on their properties and applications has progressed~\cite{OAM1,OAM3,KWZZ,KM}. These studies have been applied to fields of particle physics such as understanding the spin structure of nucleons and the separation problem of photon angular momentum in quantum electrodynamics~\cite{OAM1}.

Recently, the Curtright-Zachos (CZ) algebra~\cite{CZ,FFZ}, which is closely related to MT operators, has been revisited from a new perspective~\cite{AS4,AS3,AS2}. The CZ algebra is defined as a Hom-Lie deformation~\cite{Hom1,Hom2,Hom3} of the Virasoro algebra~\cite{AS,NQ}, and its generators can be represented using the same operational structure as MT operators. 
In its physical interpretation, not only is its relationship with noncommutative geometry on quantum planes pointed out~\cite{superCZ,superCZ2}, but it is also expected to be connected to various problems in condensed matter physics, such as the description of edge states in quantum Hall effects through $W_\infty$ structures~\cite{cft3,Winf} and electron dynamics in the strong magnetic field limit~\cite{GMP,CTZ,CTZ2}. However, the representations of CZ algebra generators typically use the conventional gauge-dependent MT operators, leaving the problem of gauge dependence unresolved.

Based on these concerns, this paper aims to introduce new gauge-invariant magnetic translation operators (generalized magnetic translations; GMT) and reconstruct CZ generators in a gauge-invariant form. A gauge-invariant formulation not only deepens our understanding of the structure of CZ algebra and its physical interpretation but also aims to provide new perspectives for research in related fields such as quantum Hall effect, noncommutative geometry, quantum groups, and conformal field theory. This new formulation has the following three significant aspects:

1. Rigorous definition of gauge-invariant magnetic translation operators (GMT): We define GMTs that eliminate the gauge dependence of conventional MT operators, allowing for clear physical interpretation. We derive the algebraic properties of GMTs (commutation relations, composition rules, etc.) and clarify their mathematical structure.

2. Construction of gauge-invariant representations of CZ generators using GMTs and gauge-invariant $\mathfrak{so}(3,1)$ algebra: The gauge-invariant $\mathfrak{so}(3,1)$ algebra describes fundamental symmetries in the Landau problem, including dilation and rotation. By incorporating the generators of this $\mathfrak{so}(3,1)$ algebra into GMTs, we represent CZ generators in a gauge-invariant form. This representation plays an important role in clarifying the physical picture of CZ algebra and expanding its range of applications. In our approach, the duality between dilatation and rotation is the key to constructing gauge-invariant representations.

3. Rigorous mathematical formulation: Formulate the definitions of MT operators and CZ generators, as well as the relationships between them, in a mathematically rigorous form. In particular, we guarantee the consistency and generality of the theory by discussing Reverting transformation (Appendix~\ref{sec:Landau}) and gauge invariance in detail.

\subsection{Settings and contents}

We consider a single-electron system confined to the $xy$ plane. Let us assume a magnetic field $\bm{B}=(0,0,B)$ perpendicular to the $xy$ plane. We denote the cyclotron frequency by $\omega$, the magnetic length by $l_B$, the cyclotron center (guiding center) by $\bm{\beta}$, and the flux quantum by $\phi_0$. (For specific definitions of each, see Appendix~\ref{sec:Landau})
\beq
H=\frac{1}{2m} ({\bm p} +\frac{e}{c}\bm{A})^2 = \frac{1}{2m}\bm{\pi}^2  \label{ham}
\eeq
\beq
p_i= -i\hbar\partial_i\,,\quad
\pi_i = -i\hbar\partial_i  + \frac{e}{c}A_i\,
\eeq
The system we are considering is a two-dimensional Euclidean plane with broken translational symmetry, and magnetic translation (MT) operations commute with the Hamiltonian instead of ordinary translations. Just as the free electron Hamiltonian becomes \eqref{ham} by replacing $\bm{p}$ with the covariant derivative $\bm{\pi}$, in the case of symmetric gauge, the element of the translation group $\exp{\frac{i}{\hbar}R\cdot p}$ is replaced by the element of the MT group $\exp{\frac{i}{l_B^2}R\cdot\pi^c}$. (In addition to the substitution $\bm{p}\ra\bm{\pi}^c$, the fundamental length changes from $\hbar\ra\l_B^2$. See Section~\ref{sec:sym_g}.)

While all functions of $p_i$ can be written as gauge-invariant quantities with this substitution, this does not necessarily preserve symmetries other than gauge invariance. As reviewed in Section~\ref{sec:magTr}, the algebraic structure of MT depends on gauge choice. Moreover, this substitution yields MT operators only in the symmetric gauge, not in general gauges. As an example of broken symmetry, the dilatation $D$ and special conformal transformations $K_i$ ($i=1,2$), which are generators of $\mathfrak{so}(3,1)$, also require modifications to their substitutions. To maintain gauge invariance, the simple substitution $p_i\ra\pi_i$ must be modified to a magnetically extended substitution. These details are explained in Section~\ref{sec:confg}.

This paper aims to define $\mathfrak{so}(3,1)$ generators and MT operators in a form where their gauge invariance is clearly recognizable. In Section~\ref{sec:confg}, using the reverting transformation technique starting from the symmetric gauge, we construct gauge-invariant $\mathfrak{so}(3,1)$ generators and discuss the duality between dilatation and rotation. The reverting transformation technique is a different approach from the methods that derived OAM~\cite{OAM1}-\cite{KM}, and while it reproduces the pseudo-angular momentum, it leads to different linear combination coefficients for pseudo-momentum.

In Section~\ref{sec:magTr}, after confirming that the algebraic relations of MT such as exchange and composition depend on gauge choice for several commonly used gauges, we present algebraic relation formulas without making gauge choices by using parametric gauge representations. Since this corresponds to gauge independence limited to a narrow range, we consider the definition of gauge-invariant MT in Section~\ref{sec:newTR} and define generalized GMT operators to handle a wider range.

This paper further focuses on the mathematical structure of GMT, specifically on the CZ algebra that can be expressed using MT operators. The generators of CZ algebra have come to play an important role in discussing the properties of CZ algebra as they can be written using MT operators~\cite{AS4}-\cite{AS2}, and there are also Lie-algebra type Virasoro algebra deformations similar to CZ algebra that can be represented by MT operators~\cite{JS,KS}. If these were gauge-dependent, then the structure constants of these algebras would be gauge-dependent, causing the inconvenience that their results and physical applications~\cite{qstr} would depend on gauge. To address these concerns, we clarify the relationship between GMT and CZ generators in Section~\ref{sec:cz}. We explain that CZ generators are based on GMT constructed through gauge-invariant dilatation and rotation, which are in a dual relationship.

The structure of this paper is as follows. In Section~\ref{sec:confg}, as preparation, we discuss in detail the gauge-invariant generators of $\mathfrak{so}(3,1)$ algebra in the Landau problem. In particular, we show the duality between the dilatation and the rotation operators, laying the foundation for later discussions. In Section~\ref{sec:magTr}, we revisit the definition of conventional MT operators and their problem of gauge dependence, and present operational relation formulas independent of gauge choice. In Section~\ref{sec:newTR}, we rigorously define gauge-invariant magnetic translation operators (GMT), one of the main results of this study, and clarify their algebraic properties. In Section~\ref{sec:ex}, we present examples of GMT, using gauge-invariant $\mathfrak{so}(3,1)$ subalgebra $\mathfrak{sim}(2)$ generators. Section~\ref{sec:cz} specifically demonstrates that CZ generators can be represented by GMT using gauge-invariant $\mathfrak{sim}(2)$ generators. Based on these results, we discuss the relationship between the $\mathfrak{sim}(2)$ duality derived in Section~\ref{sec:confg} and the internal symmetry possessed by the FFZ algebra~\cite{FFZ}. In Section~\ref{sec:end}, we summarize the conclusions of this research and discuss prospects for future research. Appendix~\ref{sec:Landau} provides details about the reverting transformation, and Appendix~\ref{sec:gauge} supplements the discussion on gauge invariance.

%%%%%%%%%%%%%%%%%%%%%%%%%%%%%%%%%%%%%%%%%%%%
%    2.  Gauge invariant generators: so(3,1) conformal group
%%%%%%%%%%%%%%%%%%%%%%%%%%%%%%%%%%%%%%%%%%%%
\section{Gauge invariant $\mathfrak{so}(3,1)$ generators}\label{sec:confg}
\setcounter{equation}{0}

In this section, we present a method to derive the gauge-invariant form of operators using the transformation technique of reverting from symmetric gauge, and construct gauge-invariant $\mathfrak{so}(3,1)$ generators. This approach reproduces the known pseudo OAM~\cite{OAM3,KM}, but yields a slightly different expression for pseudo momentum~\cite{OAM3,KM}. We demonstrate that there exists a duality between pseudo dilatation and pseudo OAM. This duality plays an important role in the construction of GMT discussed in Section~\ref{sec:newTR} and beyond.

The generators of the conformal group $\mathrm{Conf}(d)$ (translation $P_\mu$, dilatation $D$, special conformal $K_\mu$, Lorentz $M_{\mu\nu}$)
\begin{align}
&p_\mu=-i\hbar\partial_\mu\,,\quad D=x_\mu p^\mu\,,\quad K_\mu=-x^2p_\mu+2x_\mu D\,, \label{PDKdef}\\
&M_{\mu\nu}=x_\mu p_\nu-x_\nu p_\mu=-i\hbar(x_\mu\partial_\nu -x_\nu\partial_\mu)
\label{Mdef}
\end{align}
satisfy the conformal algebra. This includes the Lorentz group algebra $M_{\mu\nu}$ and the Poincar\'{e} group algebra ($M_{\mu\nu},P_\mu$) as subgroups:
\begin{align}
&[D,p_\mu]=i\hbar p_\mu\,,\quad[D,K_\mu]=-i\hbar K_\mu\,,\quad
[K_\mu,p_\nu]=2i\hbar(\delta_{\mu\nu}D+ M_{\mu\nu})     \label{DPKalg}\\
&[M_{\nu\rho},X_\mu]=i\hbar\delta_{\mu\nu}X_\rho - (\nu\lra\rho)\,,\quad\quad  X_\mu=p_\mu,K_\mu \label{MXalg}\\
&[M_{\mu\nu},M_{\rho\sigma}]=i\hbar(\delta_{\mu\rho}M_{\nu\sigma}-\delta_{\nu\rho}M_{\mu\sigma})-(\rho\lra\sigma)\,.
\label{MMalg}
\end{align}
Since we are dealing with $d=2$ Euclidean space, $M_{\mu\nu}$ is the $\mathrm{SO}(2)$ rotation operator (angular momentum) $J_3=-i\hbar l_3$ ($l_i$ is defined by \eqref{defAM})
\beq
M_{12}=J_3=-i\hbar(x_1\partial_2-x_2\partial_1)=(\bm{x}\times\bm{p})_3\,,
\eeq
and the conformal algebra \eqref{DPKalg}-\eqref{MMalg} constitutes $\mathfrak{so}(3,1)$.
Eq. \eqref{MMalg} becomes the trivial relation $[J_3,J_3]=0$ and will not be considered hereafter. The subgroup excluding $K_\mu$ forms the similarity transformation algebra $\mathfrak{sim}(2)$.

We now consider how these generators should be modified when a magnetic field is switched on. First, $J_3$ is a conserved quantity that commutes with the Hamiltonian, however $(\bm{x}\times\bm{\pi})_3$, obtained by the substitution $p_\mu\ra\pi_\mu$, does not commute with the Hamiltonian. Hence, we introduce a correction term to define a new invariant $\mathcal{J}_3$ (known as pseudo OAM~\cite{OAM3,KM}):
\beq
\mathcal{J}_3:=(\bm{x}\times\bm{\pi})_3-\frac{\hbar}{2l_B^2}\bm{x}^2\,. \label{J3cal_sec}
\eeq
This commutes with the Hamiltonian and becomes $\mathcal{J}_3=J_3$ in the symmetric gauge (see \eqref{J3cal}-\eqref{calJ3pi}). It can be obtained by first finding the harmonic oscillator representation that becomes gauge-invariant in the symmetric gauge, and then rewriting it in terms of $(\bm{x}, \bm{\pi})$ based on the gauge-independent definition of the oscillators.

Next, let us consider the dilatation $D$. If we attempt the gauge-invariant substitutions $J_3\ra\mathcal{J}_3$ and $p_\mu\ra\pi_\mu$ in the first equation of \eqref{DPKalg} and in \eqref{MXalg}, we obtain
\begin{align}
&[\mathcal{D}, \pi_\mu]=i\hbar\pi_\mu\,,  \quad\, \mathcal{D}=x_\mu\pi^\mu  \label{wrong1} \\
&[\mathcal{J}_3, \pi_\mu]=i\hbar\eps^{\mu\nu}\pi_\nu. \label{wrong2}
\end{align}
However, those obtained by the gauge-invariant prescription in Appendix~\ref{sec:Landau} are given by \eqref{calDpi} and \eqref{calJ3pi}. While \eqref{wrong2} is correct, expanding the $\beta$ term on the RHS of \eqref{calDpi} in terms of $\pi$ and $x$ yields:
\beq
[\mathcal{D},\pi_\mu]=i\hbar\pi_\mu+i\frac{\hbar^2}{l_B^2}\eps^{\mu\nu}x_\nu\,,     \label{inhomo}
\eeq
which shows the emergence of an inhomogeneous term on the right-hand side, making \eqref{wrong1} inconsistent. Moreover, as evident from \eqref{J3cal_sec}, the substitution $J_3\ra\mathcal{J}_3$ itself is not consistent with $p_\mu\ra\pi_\mu$.

In fact, such simple substitutions are inadequate, and the correct substitution to satisfy all of \eqref{DPKalg}-\eqref{MMalg} is:
\beq
p_\mu \ra P_\mu=\pi_\mu+\frac{\hbar}{2l_B^2}\eps^{\mu\nu}x_\nu   \label{covP}
\eeq
and all operators containing $p_\mu$ should be replaced with those using $p_\mu\ra P_\mu$. That is,
\beq
\mathcal{D}=x_\mu P^\mu\,,\quad \mathcal{K}_\mu=-x^2P_\mu+2x_\mu \mathcal{D}\,,\quad
\mathcal{M}_{\mu\nu}=x_\mu P_\nu-x_\nu P_\mu\,. \label{replaceP}
\eeq
For these substituted operators, \eqref{DPKalg}-\eqref{MMalg} become:
\begin{align}
&[\mathcal{D},P_\mu]=i\hbar P_\mu\,,\quad[\mathcal{D},\mathcal{K}_\mu]=-i\hbar \mathcal{K}_\mu\,,\quad
[\mathcal{K}_\mu,P_\nu]=2i\hbar(\delta_{\mu\nu}\mathcal{D}+ \eps^{\mu\nu}\mathcal{J}_3)     \label{DPKalg2}\\
&[\mathcal{J}_3,X_\mu]=i\hbar\eps^{\mu\nu}X_\nu\,,\quad\quad  X_\mu=P_\mu,\mathcal{K}_\mu \label{MXalg2}\\
&[\mathcal{J}_3,\mathcal{J}_3]=0\,,\quad \mathcal{J}_3=\mathcal{M}_{12}\,.
\end{align}
It should be noted that the second term in \eqref{covP} does not affect $\mathcal{D}$, and $\mathcal{D}$ and $\mathcal{J}_3$ are the same as those discussed in Appendix~\ref{sec:Landau}. It is worth mentioning that doubling the second term in \eqref{covP} gives the $P_\mu$ that coincides with what is known as pseudo momentum~\cite{OAM3,KM}. Following this nomenclature, $\mathcal{D}$ and $\mathcal{K}_\mu$ can be called pseudo dilatation and pseudo conformal transformation, respectively.

The commutation relations with $x_\mu$ and other commutative relations also hold as in the free electron system:
\beq
[P_\mu,P_\nu]=0\,,\quad [P_\mu,x_\nu]=-i\hbar\delta_{\mu\nu}\,,\quad
[\mathcal{D},x_\mu]=-i\hbar x_\mu\,\quad
\eeq
\beq
[\mathcal{K}_\mu,\mathcal{K}_\nu]=0\,,\quad
[\mathcal{J}_3,\mathcal{D}]=0\,,\quad [\mathcal{J}_3,x_\mu]=i\hbar\eps^{\mu\nu}x_\nu\,.
\eeq

%%%%%%%%%%%%%%%%%%%%%%%%%%%%%%%%%%
%  2.1  prescription for gauge invariance
%%%%%%%%%%%%%%%%%%%%%%%%%%%%%%%%%
\subsection{Prescription}\label{sec:method}
\indent

Here, we explain a systematic method for deriving $P$, $\mathcal{D}$, $\mathcal{J}_3$, and $\mathcal{K}_\mu$. In Appendix~\ref{sec:Landau}, $\mathcal{D}$ and $\mathcal{J}_3$ are derived heuristically by considering the characteristics of the symmetric gauge.

Organizing that method, we prepare physical quantities $Q$ that can be explicitly written in a gauge-independent form. Let $Q_{sym}$ be the corresponding physical quantities in the symmetric gauge. If an operator $G$ can be expressed as a function of $Q_{sym}$, $G=F(Q_{sym})$, then in any gauge, it should be expressible as $\mathcal{G}=F(Q)$, and we utilize this fact. In the symmetric gauge, holding $\mathcal{J}_3=J_3$, $\mathcal{D}=D$, and $\mathcal{K}_\mu=K_\mu$, we have the advantage of starting from the expressions for a free electron without a magnetic field.

For the convenience of the duality $\mathcal{D}\lra\mathcal{J}_3$ explained in the next section, we use the complex representation of the symmetric gauge. First, taking the complex representation of $K_\mu$:
\beq
K_\pm=\frac{1}{2}(K_1\pm iK_2)\,,\quad z=x+iy\,,
\eeq
the holomorphic and anti-holomorphic parts can be separated as:
\beq
K_+=-i\hbar z^2\partial\,,\quad K_-=-i\hbar \bar{z}^2\bar{\partial} \,. \label{Kpm}
\eeq
Rearranging these, we obtain the complex differential representation of the first and second components of $K_\mu$:
\beq
K_1=-i\frac{\hbar}{2}(z^2\partial+\bar{z}^2\bar{\partial})\,,\quad K_2=-\frac{\hbar}{2}(z^2\partial-\bar{z}^2\bar{\partial})\,. \label{Kz}
\eeq
The commutation relations of $K_\pm$ with $D$, $J_3$, and $p_\mu$ can be expressed using the complex representation of $p_\mu$:
\beq
 p_\pm=\frac{1}{2}(p_1\mp ip_2)\,,\quad p_+=-i\hbar\partial=-i\hbar\partial_z\,,\quad 
p_-=-i\hbar\bar{\partial}=i\hbar\partial^\dagger\,,
\eeq
as follows:
\beq
[K_+,K_-]=0\,,\quad [J_3,K_\pm]=\pm\hbar K_\pm\,,\quad [D,K_\pm]=-i\hbar K_\pm  \label{KJDalg}
\eeq
\beq
[K_\pm,p_\mp]=0\,,\quad\,
[K_\pm,p_\pm]= i\hbar(D\mp i J_3)\,.  \label{KPalg}
\eeq

Returning to the two-dimensional electron system in a magnetic field, in the symmetric gauge, the complex coordinates $z,\bar{z}$ and their differential operators $p_\pm$ can be expressed in terms of two commuting harmonic oscillators $a,b$:
\begin{align}
&z=\sqrt{2}l_B(a^\dagger+b)\,,\quad \bar{z}=\sqrt{2}l_B(a+b^\dagger)\,, \label{z2a} \\
&\partial=\frac{\sqrt{2}}{4l_B}(a-b^\dagger)\,,\quad \bar{\partial}=\frac{\sqrt{2}}{4l_B}(b-a^\dagger)\,. \label{del2a}
\end{align}
Substituting these into the complex differential representation of $K_\mu$ \eqref{Kz}, we obtain the harmonic oscillator representation, which is gauge-invariant. First, let us confirm that this prescription works for $D$ and $J_3$. The substitution results in:
\begin{align}
&D=-i\hbar(z\partial+\bar{z}\bar{\partial})=i\hbar(a^\dagger b^\dagger-ab+1)\,, \label{Dharm} \\
&J_3=-i\hbar l_3=\hbar(z\partial-\bar{z}\bar{\partial})=\hbar(a^\dagger a-b^\dagger b)\,, 
\end{align}
and these right-hand sides indeed match the gauge-invariant expressions \eqref{Dcal} and \eqref{J3cal}.

These harmonic oscillators, originally being linear combinations of $x_\mu$ and $\pi_\mu$, 
are themselves gauge-invariant. It follows that the polynomials $\mathcal{D}$ and $\mathcal{J}_3$, being composed of these ocillators, are also gauge-invariant. 
In fact, \eqref{covP} is derived based on this prescription: $P_\mu$ is obtained by expressing $p_\mu$ in terms of complex derivatives, transforming it into harmonic oscillator variables, and then rewriting the result using $\pi_\mu$. 
$P_\mu$ is thus gauge-invariant, and the substitution $p_\mu\ra P_\mu$ preserves the conformal algebra.

Similarly, from the complex representation of $K_\mu$ \eqref{Kpm}, we obtain the harmonic oscillator representation of $K_\pm$, denoted as $\mathcal{K}_\pm$:
\beq
\mathcal{K}_+=-i\hbar\frac{l_B}{\sqrt{2}}(a^\dagger+b)^2(a-b^\dagger)\,,\quad 
\mathcal{K}_-=-i\hbar\frac{l_B}{\sqrt{2}}(a+b^\dagger)^2(b-a^\dagger)\,,
\eeq
and the algebraic relations \eqref{KJDalg} and \eqref{KPalg} still hold for $P_\mu$, $\mathcal{K}_\pm$, $\mathcal{D}$, and $\mathcal{J}_3$ in the same forms.

%%%%%%%%%%%%%%%%%%%%%%%%%%%%%%%5
%   2.2   (D, J3)  exchange symmetry　　
%%%%%%%%%%%%%%%%%%%%%%%%%%%%%%%%
\subsection{$\mathfrak{sim}(2)$ duality between $\mathcal{D}$ and $\mathcal{J}_3$}\label{sec:dual}
\indent

Continuing to use the complex differential representation in the symmetric gauge, we decompose $\mathcal{D}$ and define $\mathcal{D}_\pm$ as
\beq
\mathcal{D}_+=zp_+=\frac{1}{2}(D-iJ_3)=-i\hbar z\partial\,,\quad 
\mathcal{D}_-=\bar{z}p_-=\frac{1}{2}(D+iJ_3)=-i\hbar \bar{z}\bar{\partial}\,.  \label{calD+D-}
\eeq
With this decomposition, we have
\beq
\mathcal{D}=\mathcal{D}_++\mathcal{D}_-\,,
\eeq
and the commutation relations \eqref{KPalg} take the form
\beq
[K_\pm,p_\pm]=2i\hbar\mathcal{D}_\pm\,.
\eeq
$\mathcal{D}_\pm$ are commuting scale transformations accompanied by imaginary rotation components, representing scaling operations in the radial direction while rotating in opposite directions. Using the polar coordinate representation $z=r\exp{i\theta}$, we can see that $\mathcal{D}$ and $\mathcal{J}_3$ correspond to radial scaling and rotation operators, respectively:
\begin{align}
&\mathcal{D}=\mathcal{D}_++\mathcal{D}_-=-i\hbar(z\partial+\bar{z}\bar{\partial})=-i\hbar r\partial_r\,,  \label{Pr}\\
&\mathcal{J}_3=i(\mathcal{D}_+-\mathcal{D}_-)=\hbar(z\partial-\bar{z}\bar{\partial})=-i\hbar\partial_\theta\,.  \label{Ptheta}
\end{align}
Since the relationship between $D_\pm$ and $J_3$ holds independently of the choice of  gauge, the following commutation relation is valid (this can be verified using \eqref{z2a} and \eqref{del2a} to express $\mathcal{D}_\pm$ in terms of harmonic oscillators):
\beq
[\mathcal{K}_\pm, P_\pm]=2i\hbar\mathcal{D}_\pm\,,
\eeq
where
\beq
\mathcal{D}_+=\frac{1}{2}(\mathcal{D}-i\mathcal{J}_3)\,,\quad \mathcal{D}_-=\frac{1}{2}(\mathcal{D}+i\mathcal{J}_3)\,. \label{D+-}
\eeq

Let us now focus on the exchange symmetry of operators. $\mathcal{D}_\mp$ can be obtained from $\mathcal{D}_\pm$ by interchanging $J_3$ and $D$ in \eqref{calD+D-},  multiplied by $\pm i$. Explicitly, we have
\beq
\mathcal{D}_-=i\mathcal{D}_+[J_3\lra D]\,,\quad  \mathcal{D}_+ =-i\mathcal{D}_-[J_3\lra D]\,.
\eeq
This reflects a duality in which the exchange of the operators $J_3 \lra D$ corresponds to the exchange $i\mathcal{D}_+ \lra \mathcal{D}_-$. Consequently, the roles of $\mathcal{D}$ and $\mathcal{J}_3$ are also interchanged:
\beq
\mathcal{D} \lra \mathcal{J}_3\,.
\eeq
This exchange is equivalent to the interchange of holomorphic and anti-holomorphic in scale transformations (accompanied by a 90-degree rotation of the plane):
\beq
iz\partial \lra \bar{z}\bar{\partial}\,,
\eeq
and signifies the duality between radial scaling \eqref{Pr} and rotational operations \eqref{Ptheta}.

Note that $P_\pm$, $\mathcal{D}_\pm$, and $K_\pm$ are separated into (commuting) holomorphic and anti-holomorphic parts, as seen in \eqref{Kpm} and \eqref{calD+D-}. The holomorphic set $\{P_+, \mathcal{D}_+, K_+\}$ forms the operators $L_n = -i\hbar z^{n+1}\partial$ $(n = 0, \pm1)$, which generate the $\mathfrak{sl}(2,\mathbb{C})$ subalgebra of the Virasoro algebra. Similarly, the anti-holomorphic set $\{P_-, \mathcal{D}_-, K_-\}$ forms $\bar{L}_n$, where the scale transformation is given by $L_0 + \bar{L}_0$ and the angular momentum operator by $i(L_0 - \bar{L}_0)$.

%%%%%%%%%%%%%%%%%%%%%%%%%%5
%   2.3  dual su(2)
%%%%%%%%%%%%%%%%%%%%%%%%%%%
\subsection{Dual $\mathfrak{su}(2)$ derived from $\mathcal{D}$}
\indent

Let us now consider a different perspective from holomorphic and anti-holomorphic to examine the decomposition of \eqref{Dharm}. 
If $\mathcal{D}$ possesses a symmetry with $\mathcal{J}_3$ by role exchange, then $\mathcal{D}$ must inherently contain properties of a rotation operator. (For instance, $\mathcal{D}$ should be decomposable into non-commutative $\mathfrak{su}(2)$ operators, suggesting that the seeds of $J_3$'s role are hidden within.)

Focusing on the harmonic oscillator representation of $\mathcal{D}$, if we define $\mathscr{D}_i = (\mathscr{D}_1,\mathscr{D}_2,\mathscr{D}_3)$ and $\mathscr{D}_\pm$ as follows:
\begin{align}
&\mathscr{D}_+=i\hbar a^\dagger b^\dagger\,,\quad \mathscr{D}_-=i\hbar ab\,,\quad 
\mathcal{D}=\mathscr{D}_+-\mathscr{D}_-+i\hbar\,,  \label{scrD1}\\
&\mathscr{D}_1=\frac{1}{2}(\mathscr{D}_++\mathscr{D}_-)=\frac{i\hbar}{2}(a^\dagger b^\dagger+ab)\,,\quad
\mathscr{D}_2=\frac{1}{2i}(\mathscr{D}_+-\mathscr{D}_-)=\frac{\hbar}{2}(a^\dagger b^\dagger-ab)\,, \label{scrD2} \\
&\mathscr{D}_3=\frac{\hbar}{2}(a^\dagger a+bb^\dagger)\,,
\end{align}
then $\mathscr{D}_i$ satisfies the same algebraic relations as the angular momentum operators \eqref{JJalg1} and \eqref{JJalg2}:
\begin{align}
& [\mathscr{D}_i,\mathscr{D}_j]=i\hbar\eps^{ijk}\mathscr{D}_k\,,\quad  [\mathscr{D}_i,\mathscr{D}^2]=0\,,  \label{DDalg1}  \\
& [\mathscr{D}_3,\mathscr{D}_\pm]=\pm\hbar \mathscr{D}_\pm\,,\quad [\mathscr{D}_+,\mathscr{D}_-]=2\hbar \mathscr{D}_3\,.
\label{DDalg2}
\end{align}
Note that
\beq
\mathscr{D}^2=\sum_{i=1}^3=\mathscr{D}_3^2+\frac{1}{2}(\mathscr{D}_+\mathscr{D}_-+\mathscr{D}_-\mathscr{D}_+)\not=D^2 \,.
\eeq
From the harmonic oscillator representation, we find
\beq
H=\omega \mathscr{D}_3+\frac{\omega}{2}\mathcal{J}_3\,.
\eeq
Comparing this with the expression recalled from \eqref{HisJ3beta}:
\[
H=\omega\mathcal{J}_3+\frac{\omega\hbar}{2l_B^2}\bm{\beta}^2\,,
\]
we deduce the relation
\beq
\mathscr{D}_3=\frac{1}{2}\mathcal{J}_3+\frac{\hbar}{2l_B^2}\bm{\beta}^2\,.  \label{J2Deffect}
\eeq
Noting that $\bm{\beta}$ commutes with $H$, and using this relation together with \eqref{DDalg1} and \eqref{DDalg2}, we immediately obtain
\beq
[\mathscr{D}_3,H]=0\,,\quad[\mathscr{D}_3,\mathcal{J}_3]=0\,,\quad [\mathscr{D}_\pm,\mathcal{J}_3]=0\,,
\eeq
\beq
[H,\mathscr{D}_\pm]=\pm\hbar\omega \mathscr{D}_\pm\,,\quad [H,\mathscr{D}^2]=0\,.
\eeq
These relations indicate that $\mathscr{D}_3$ can be simultaneously diagonalized with $H$ and $\mathcal{J}_3$, and that the energy eigenvalue can be decomposed into two components associated with $\mathscr{D}_3$ and $\mathcal{J}_3$. Since $\bm{\beta}^2$ commutes with the Hamiltonian, it follows that, apart from the contribution of $\bm{\beta}^2$, $2\mathscr{D}_3$ is essentially equivalent to $\mathcal{J}_3$. Equation \eqref{J2Deffect} thus shows that the displacement of the center of rotational motion manifests as a dilatation effect, and that $\omega\mathscr{D}_3$ corresponds to the harmonic oscillator Hamiltonian with half of the angular momentum contribution removed.

For reference, we summarize the harmonic oscillator representations of $\mathscr{D}_i$ in both complex and polar coordinate systems. In the complex coordinate representation, $\mathscr{D}_1$ corresponds to the negative Laplacian (with an opposite sign compared to $\mathscr{D}_3$), while $\mathscr{D}_2$ generates radial scaling (dilatation). The operator $\mathscr{D}_3$ acts on eigenstates with constant radius in the $\pi$-$\beta$ space.
\begin{align}
\mathscr{D}_1: \quad&a^\dagger b^\dagger +ab=4\partial\bar{\partial}+\frac{|z|^2}{4l_B^4}=\Delta+\frac{r^2}{4l_B^4} \\ 
\mathscr{D}_2: \quad&ab-a^\dagger b^\dagger=z\partial+\bar{z}\bar{\partial}+1=r\partial_r+1 \\ 
\mathscr{D}_3: \quad&a^\dagger a+bb^\dagger = \frac{l_B^2}{2\hbar^2}\bm{\pi}^2+\frac{1}{l_B^2}\bm{\beta}^2 =-l_B^2\Delta+\frac{r^2}{4l_B^2}\,,
\end{align}
where
\beq
\Delta=4\partial\bar{\partial}=\frac{1}{r}\partial_r r\partial_r +\frac{1}{r^2}\partial_\theta\,.
\eeq

The momentum representation in polar coordinates is given by
\begin{align}
&rP_r=-i\hbar(1+r\partial_r)=-i\hbar(z\partial+\bar{z}\bar{\partial}+1)=i\hbar(a^\dagger b^\dagger-ab)\,, \\ 
&P_\theta=-i\hbar\partial_\theta=\hbar(z\partial-\bar{z}\bar{\partial})=\hbar(a^\dagger a -b^\dagger b)\,.
\end{align}
Comparing \eqref{scrD1} and \eqref{scrD2}, we find that these operators are related to $\mathcal{D}$ and $\mathcal{J}_3$ as
\beq
\mathcal{D}=2i\mathscr{D}_2+i\hbar =rP_r+i\hbar=-i\hbar r\partial_r\,,
\eeq
\beq
\mathcal{J}_3=-i\hbar\partial_\theta\,.
\eeq
If we map the complex plane coordinates $z=re^{i\theta}=(r,\theta)$ to the cylindrical coordinates $w=\tau+i\theta=(\tau,\theta)$, the $\mathcal{D}\leftrightarrow\mathcal{J}_3$ duality manifests as a $\tau\leftrightarrow\theta$ exchange:
\beq
\mathcal{D}=P_\tau=-i\hbar\partial_\tau\,,\quad \mathcal{J}_3=P_\theta=-i\hbar\partial_\theta\,, \label{cylinDJ3}
\eeq
where the differential forms remain the same except for the periodicity in $\theta$.
These operators form a fundamental conjugate pair that underpins the construction of GMT, as discussed in Section~\ref{sec:cz}, and play a central role in defining the CZ generators.

%%%%%%%%%%%%%%%%%%%%%%%%%%%%%%%%%%%%%%%%%%%%%%%%%%%%%%%%%%%
%         3.    Magnetic transaltions
%%%%%%%%%%%%%%%%%%%%%%%%%%%%%%%%%%%%%%%%%%%%%%%%%%%%%%%%%%%
\section{Magnetic translation (MT) operators}\label{sec:magTr}
\setcounter{equation}{0}
\indent

Magnetic translation (MT) operators are fundamental tools in the study of condensed matter physics; however, most existing discussions are formulated under a specific gauge choice. The primary objective of this section is to elucidate that the algebraic relations of MT operators such as exchange and composition are inherently dependent on the choice of gauge, and to establish the consistency of their operational relations and phase formulas through parametric connections between the Landau gauges and the symmetric gauge, without fixing any specific gauge.

This section also serves as a preliminary review to introduce notations and clarify definitions. Notations for MT operators vary widely across the literature, and even within a single reference, inconsistencies often arise — particularly regarding the signs associated with charges or magnetic fields. In order to facilitate rigorous and unambiguous developments in Section~\ref{sec:newTR} and beyond, it is essential to systematically organize these conventions and definitions, thereby avoiding unexpected inconsistencies or ambiguities in interpretation.

First, ordinary translation operators $T_{\bm R}$ satisfy the following relations:
\beq
T_{\bm R} =\Exp{\frac{i}{\hbar} \bm{R}\cdot\bm{p}} =\Exp{R_1\partial_1+R_2\partial_2}\,,
\eeq
\beq
[T_{\bm{R}_1}, T_{\bm{R}_2}]=0\,,\quad  T_{\bm{R}_1}T_{\bm{R}_2} =T_{\bm{R}_1+\bm{R}_2}\,,
\eeq
\beq
T_{\bm{R}}^\dagger =T_{\bm{R}}^{-1}\,,\quad
T_{\bm R}   \mathcal{O}(\bm{x}) T_{\bm R}^{-1}  = \mathcal{O}(\bm{x}+\bm{R})\,.
\eeq
Since a uniform magnetic field satisfies
\beq
\bm{B}=\mathrm{rot}\bm{A}(\bm{x})=\mathrm{rot}\bm{A}(\bm{x}+\bm{R})
\eeq
for any $\bm{R}$, there exists a scalar function (gauge function) $\xi$ such that
\beq
\bm{A}(\bm{x}+\bm{R})-\bm{A}(\bm{x})=\nabla \xi(\bm{x},\bm{R})\,.  \label{gaugef}
\eeq
Using the basis vectors $\bm{e}_1$ and $\bm{e}_2$ of a two-dimensional lattice, we can express the lattice points as
\beq
\bm{R}=n\bm{e}_1+m\bm{e}_2\,, \quad\mbox{or equivalently,}\quad (R_x,R_y)=(n,m)\,, \label{Rbases}
\eeq
where $n,m\in\mathbb{Z}$. (Strictly speaking, $\bm{R}=(na,ma)$ with $a$ representing the lattice constant. For simplicity, we set $a=1$ here, and will restore $a$ explicitly when necessary.) Assuming linearity of the gauge function, we have
\begin{align}
&\xi(\bm{x},\bm{R})=n\xi(\bm{x},\bm{e}_1) +m\xi(\bm{x},\bm{e}_2) \,, \label{e1e2} \\
&\xi(\bm{x}_1+\bm{x}_2,\bm{R})=\xi(\bm{x}_1,\bm{R}) +\xi(\bm{x}_2,\bm{R})\,. \label{x1x2}
\end{align}
From these properties, it follows that
\beq
\bm{A}(\bm{x}-\bm{R})-\bm{A}(\bm{x})=-\nabla \xi(\bm{x},\bm{R})\,. \label{gaugef2}
\eeq

The MT operator $\Tm{R}$ is defined by
\beq
\Tm{R}= \Exp{2\pi i\frac{\xi}{\phi_0}} T_{\bm R}\,,\quad  \phi_0=\frac{hc}{e}\,. \label{TmR}
\eeq
Using \eqref{gaugef} and \eqref{gaugef2}, we can show that
\begin{eqnarray}
\Tm{R} \bm{\pi} \Tm{R}^{-1} &=& \Exp{2\pi i\frac{\xi}{\phi_0}} T_{\bm R} \left(\bm{p}+\frac{e}{c}\bm{A}\right)
T_{\bm R}^{-1}\Exp{-2\pi i\frac{\xi}{\phi_0}}  \nn \\
&=& \bm{p}-\frac{e}{c}\nabla\xi + \frac{e}{c}\bm{A}(\bm{x}+\bm{R}) \nn \\
&=& \bm{p}+\frac{e}{c}\bm{A}(\bm{x})\,,
\end{eqnarray}
which demonstrates that $\Tm{R}$ commutes with the Hamiltonian \eqref{ham} independently of the gauge choice:
\beq
\Tm{R} \bm{\pi} \Tm{R}^{-1} = \bm{\pi}\,,\quad \mathrm{i.e.}\quad [\Tm{R},\bm{\pi}]=0
\quad \ra\quad [H,\Tm{R}]=0\,.
\eeq

Similarly, the charge-conjugate MT operator is defined as
\beq
\Tm{R}^\ast= \Exp{-2\pi i\frac{\xi}{\phi_0}} T_{\bm R}\,,
\eeq
and satisfies
\beq
\Tm{R}^\ast \bm{\pi}^c {\Tm{R}^\ast}^{-1} = \bm{p}-\frac{e}{c}\bm{A}(\bm{x})\,,
\eeq
\beq
\Tm{R}^\ast \bm{\pi}^c {\Tm{R}^\ast}^{-1} = \bm{\pi}^c\,,
\quad \mathrm{i.e.}\quad [\Tm{R}^\ast,\bm{\pi}^c]=0\,,
\quad [\pi_i, \pi^c_j]=0
\quad \ra\quad [H,\Tm{R}^\ast]=0\,.
\eeq

It is natural to expect that $\Tm{R}$ can be expressed in terms of $\bm{\beta}$ (or $b$, $b^\dagger$), since these operators commute with the Hamiltonian $H$. This expectation can be verified by choosing specific gauge functions, such as the Landau gauge or the symmetric gauge. In some cases, $\Tm{R}$ is expressed in terms of $\bm{\pi}$ (as in the symmetric gauge), but this is not a preferable representation, because $\bm{\pi}$ does not commute with $H$. (Strictly speaking, it is $\bm{\pi}^c$ that should be considered, but for our purposes, this is also undesirable; see \eqref{symTRbeta} for details.)

To explicitly confirm these points, in the following subsections we will compare two specific gauge choices and then extend the discussion to general gauges.

%%%%%%%%%%%%%%%%%%%%%%%%%%%%%%%
%  3.1  Example 1 Landau Gauge
%%%%%%%%%%%%%%%%%%%%%%%%%%%%%%%
\subsection{Landau gauge}\label{sec:landau_g}

For the Landau gauge, there are two cases: $\bm{A}=(0,Bx,0)$ (standard/first Landau gauge) and $\bm{A}=(-By,0,0)$ (anti/second Landau gauge), both giving $\bm{B}=(0,0,B)$. If we consider a weighted average of these with an arbitrary real number $\alpha$, we have
\beq
\bm{A}=(-\alpha By, (1-\alpha)Bx,0)\,,
\eeq
where the former corresponds to $\alpha=0$ and the latter to $\alpha=1$. The symmetric gauge corresponds to $\alpha=\frac{1}{2}$. 
Introducing the following parametric representation
\begin{align}
&\xi(\bm{x},\bm{R})=\alpha \xi_1(\bm{x},\bm{R})+(1-\alpha)\xi_0(\bm{x},\bm{R})  \label{xialpha}  \\
&\xi_1(\bm{x},\bm{R})=-B R_y x \,,\quad \xi_0(\bm{x},\bm{R})=B R_x y \,,  \label{Lgauge}
\end{align}
we have the commuation relation
\beq
[\xi,\bm{R}\cdot\bm{p}]=i\hbar R_xR_y B(1-2\alpha) \,. \label{xiRp}
\eeq

In this paper, we set $\alpha=1$ for the Landau gauge, choosing gauge function $\xi=\xi_1$. From \eqref{TmR} and \eqref{xiRp}, 
%\beq
%[\xi_1,\bm{R}\cdot\bm{p}]=-i\hbar R_xR_y B\,,
%\eeq
we obtain
\begin{eqnarray}
\Tm{R} &=& \Exp{-\frac{ie}{\hbar c} BR_y x } T_R \\
&=&\Exp{i\pi\frac{B}{\phi_0}R_xR_y}\Exp{ \frac{i}{l_B^2}\bm{R}\times\bm{\beta}\cdot\bm{n}  } \,,\quad \bm{n}=(0,0,1) \,.\label{LandTR}
\end{eqnarray}
Here, it is clear that $\Tm{R}$ commutes with $H$ since it is expressed as a function of $\bm{\beta}$. In the complex representation, the expressions become
\beq
\varepsilon=\frac{1}{\sqrt{2}l_B^2}(R_x+iR_y)\,,\quad \bar{\varepsilon}=\frac{1}{\sqrt{2}l_B^2}(R_x-iR_y)\,, 
\eeq
\beq
\Tm{R}=\Exp{i\pi\frac{B}{\phi_0}R_xR_y }\Exp{ \bar{\varepsilon}b - \varepsilon b^\dagger}\,.
\eeq
Let us now verify the exchange and composition rules:
\begin{align}
&\Tm{R_1}\Tm{R_2} =\Exp{-2\pi i\frac{BS}{\phi_0}} \Tm{R_2}\Tm{R_1} \,, \label{TRTRgauge}\\
&\Tm{R_1}\Tm{R_2} =\Exp{ -\frac{ie}{\hbar c} B R^2_y R^1_x} \Tm{R_1+R_2}\,. \label{TRTRldu}
\end{align}
where $S$ denotes the area of the parallelogram spanned by vectors $\bm{R}_1$ and $\bm{R}_2$,
\beq
S=\bm{n}\cdot(\bm{R}_1\times\bm{R}_2)\,.
\eeq
It is important to note that in \eqref{TRTRldu}, the term $R^2_y R^1_x$ does not directly correspond to the area $S$. This contrasts with the symmetric gauge, where the corresponding term matches $S$ exactly.

%%%%%%%%%%%%%%%%%%%%%%%%%%%%%%%
%  3.2  Example 2   Symmetric Gauge
%%%%%%%%%%%%%%%%%%%%%%%%%%%%%%%
\subsection{Symmetric gauge}\label{sec:sym_g}

In the case of symmetric gauge, we choose $\alpha=\frac{1}{2}$, and $\bm{A}=\frac{1}{2}\bm{B}\times\bm{x}=(-\frac{B}{2}y,\frac{B}{2}x,0)$, with the corresponding gauge function
\beq
\xi(\bm{x},\bm{R})=\frac{1}{2}\bm{B}\times\bm{R}\cdot\bm{x}\,.  \label{Sgauge}
\eeq
From the definition of $\Tm{R}$ in \eqref{TmR} and commutation relation \eqref{xiRp}
\beq
\xi(\bm{x},\bm{R})=-\frac{1}{2}\bm{B}\times\bm{x}\cdot\bm{R}=-\bm{A}\cdot\bm{R}\,, \quad
[\xi,\bm{R}\cdot\bm{p}]=0\,,
\eeq
we have
\beq
\Tm{R}=\Exp{\frac{ie}{2\hbar c} \bm{B}\times \bm{R}\cdot \bm{x} } T_R 
=\Exp{\frac{i}{\hbar}\bm{R}\cdot\bm{\pi}^c}   \label{symTR}
\eeq
and
\beq
\Tm{R}^\ast=\Exp{-\frac{ie}{2\hbar c} \bm{B}\times \bm{R}\cdot \bm{x} } T_R 
=\Exp{\frac{i}{\hbar}\bm{R}\cdot\bm{\pi}}\,.
\eeq
Note that it is the charge-conjugate $\Tm{R}^\ast$ that is expressed in terms of $\bm{\pi}$, not $\Tm{R}$. (The roles can be exchanged by reversing the charge, by taking the opposite sign of $\xi$, or by changing the order of $T_R$ and the phase factor in the definition of $\Tm{R}$. Due to these possibilities, these notations are often confused or ambiguous in the literature.) Additionally, in the symmetric gauge, $\bm{\pi}^c$ can be expressed in terms of $\bm{\beta}$ with exchanged axes, that is, $\pi_i^c=-\frac{\hbar}{l_B^2}\eps^{ij}\beta_j$ (see \eqref{beta2pi}). Therefore, the structure of the exponent part of $\Tm{R}$ being given by the normal component of $\bm{R}\times\bm{\beta}$ is, apart from the additional phase factor, exactly the same as in the Landau gauge \eqref{LandTR} (see \eqref{symTRbeta}):
\beq
\Tm{R}=\Exp{\frac{i}{l_B^2}\bm{R}\times\bm{\beta}\cdot\bm{n}}\,.  \label{TRsymbeta}
\eeq

The commutation rule takes the same form as the Landau gauge \eqref{TRTRgauge}, however the composition rule differs from \eqref{TRTRldu}:
\beq
\Tm{R_1}\Tm{R_2} = \Exp{-\pi i \frac{BS}{\phi_0}} \Tm{R_1+R_2}\,.  \label{TRTRsym} 
\eeq
The form of the MT commutator yields from \eqref{TRTRsym}:
\beq
[\Tm{R_1},\Tm{R_2}]=\left\{ \Exp{-\pi i\frac{BS}{\phi_0}} -\Exp{\pi i\frac{BS}{\phi_0}}   \right\} \Tm{R_1+R_2}\,.\label{comsym}
\eeq

As will be discussed in Section~\ref{sec:newTR}, the Landau gauge expression is gauge-dependent. In that section, we will construct gauge-invariant magnetic translation (MT) operators that obey the same composition and exchange relations irrespective of the chosen gauge. It will also be shown that these gauge-invariant operators take the same algebraic form as those obtained in the symmetric gauge presented here.

%\vskip\baselineskip
%%%%%%%%%%%%%%%%%%%%%%%%%%%%%%%%%
%   3.3  any gauge & sign reflections
%%%%%%%%%%%%%%%%%%%%%%%%%%%%%%%%%%
\subsection{ Arbitrary gauge and remarks}\label{sec:any_g}
\indent

In this subsection, we summarize the commutation relations, composition rules, and circulation identities that hold in an arbitrary gauge, without fixing the value of $\alpha$. Here, “arbitrary” refers specifically to the class of gauges parameterized by $\alpha$ as defined in \eqref{xialpha}. We also include a remark on the consistency of phase conventions, which often vary across different references.

Previously, we considered the cases $\alpha=1$ and $\alpha=1/2$ separately. However, substituting the general gauge function \eqref{xialpha} directly into the definition of the MT operator \eqref{TmR}, we obtain:
\begin{eqnarray}
\Tm{R} &=& \Exp{\frac{ie}{\hbar c} (\alpha\xi_1+(1-\alpha)\xi_0) } T_R \nn \\
&=& \Exp{i\pi\frac{B}{\phi_0}R_xR_y(1-2\alpha)} \Exp{ \frac{i}{l_B^2} (\bm{R}\times\bm{\beta})\cdot\bm{n} } \,, \label{anyTmR}
\end{eqnarray}
demonstrating that within the gauge class defined by \eqref{xialpha}, $\Tm{R}$ can always be expressed in terms of $\bm{\beta}$. This expression naturally reproduces the Landau gauge result \eqref{LandTR} for $\alpha=1$ and the symmetric gauge result \eqref{symTR} for $\alpha=1/2$. Moreover, we expect that the composition laws, which appear to differ in form between \eqref{TRTRldu} and \eqref{TRTRsym}, can also be unified into a single expression valid for arbitrary $\alpha$.

%\vskip\baselineskip
While both gauge choices satisfy the same commutation rule \eqref{TRTRgauge}, many literature conventionally express the phase with a positive sign. Therefore, it is beneficial to clarify the consistency with conventional notation. For this reason, from here until the end of Section~\ref{sec:any_g} only, we redefine $\xi$ with the opposite sign in selections \eqref{Lgauge} or \eqref{Sgauge} (corresponding to the reversal of the sign of $B$ or $e$), and express the commutation rule as
\beq
\Tm{R_1}\Tm{R_2} = \Exp{2\pi i \frac{BS}{\phi_0}} \Tm{R_2}\Tm{R_1}\,.  \label{newTRTR}
\eeq
Also, by the linearity \eqref{e1e2} and \eqref{x1x2}, the seletions \eqref{Lgauge} and \eqref{Sgauge} can be summarized as
\begin{align}
&\xi(\bm{e_1},\bm{e_1})=\xi(\bm{e_2},\bm{e_2})=0\,,  \label{xi1122} \\
&\xi(\bm{e_1},\bm{e_2})=a^2\alpha B\,, \quad \xi(\bm{e_2},\bm{e_1})=a^2(\alpha-1)B,   \label{Lxi12} 
\end{align}
(where the sign has already been reversed. $a$ represents the unit length, corresponding to the absolute values of the basic vectors $\bm{e}_i$ seen in \eqref{Rbases}). We apply these equations when making gauge choices in the following discussion.

This sign reversibility stems from the fact that the choice of coordinate system is not unique. When viewing the $x$-$y$ plane from the opposite side and using
\beq
(x,y)\ra(x',y')=(y,x)\,,\quad B\ra B'=-B
\eeq
the same phenomenon is described in the $x'$-$y'$ coordinate system with the opposite sign of the magnetic field. Such transformations correspond to different ways of flipping the surface, and there are two more possibilities:
\beq
(x,y)\ra(-x,y)\,,\quad \quad (x,y)\ra(x,-y)\,.
\eeq
These cause the direction of motion to appear reversed, which can also correspond to taking the opposite sign of the current (charge $e$).

In the following, we present more general formulation without using the specific expression \eqref{xialpha} for $\xi$. Keeping the general form of \eqref{TmR} without specifying the gauge, the commutation relation between MT operators becomes
\beq
\Tm{R_1}\Tm{R_2}=\Exp{ \frac{2\pi i}{\phi_0}\Xi(\bm{R_1},\bm{R_2}) }\Tm{R_2}\Tm{R_1} \,,\label{TRTRXi}
\eeq
where the phase factor $\Xi$ is defined by
\begin{align}
&\Xi(\bm{R_1},\bm{R_2})=\Xi_1(\bm{R_1},\bm{R_2})-\Xi_2(\bm{R_1},\bm{R_2}) \\
&\Xi_1(\bm{R_1},\bm{R_2})=\xi(\bm{x},\bm{R_1})-\xi(\bm{x+R_2},\bm{R_1}) \\
&\Xi_2(\bm{R_1},\bm{R_2})=\xi(\bm{x},\bm{R_2})-\xi(\bm{x+R_1},\bm{R_2}) =\Xi_1(\bm{R_2},\bm{R_1})\,.
\end{align}
Using the linearity property \eqref{x1x2}, we obtain
\beq
\Xi_1=-\xi(\bm{R_2},\bm{R_1})\,,\quad \Xi_2=-\xi(\bm{R_1},\bm{R_2})\,.
\eeq
Further, substituting $\bm{R}_i=n_i\bm{e}_1 + m_i\bm{e}_2$, we get
\begin{eqnarray}
\Xi&=&\xi(\bm{R_1},\bm{R_2})-\xi(\bm{R_2},\bm{R_1})  \label{XI} \\ 
&=&(n_1m_2-n_2m_1)\{\xi(\bm{e}_1,\bm{e}_2)-\xi(\bm{e}_2,\bm{e}_1)\} \nn \\
&=& \frac{S}{a^2}\{\xi(\bm{e}_1,\bm{e}_2)-\xi(\bm{e}_2,\bm{e}_1)\}\,.  \nn  
\end{eqnarray}
The circulation operation around a parallelogram follows from \eqref{TRTRXi}:
\beq
\Tm{R_1}^{-1}\Tm{R_2}^{-1}\Tm{R_1}^{}\Tm{R_2}^{}=e^{2\pi i \phi}\,,\quad \phi=\frac{\Xi}{\phi_0}\,.\label{4angle}
\eeq
Using the specific gauge expression \eqref{Lxi12}, for any $\alpha$, we have
\beq
\xi(\bm{e}_1,\bm{e}_2)-\xi(\bm{e}_2,\bm{e}_1)=Ba^2\,,
\eeq
leading to the well-known dimensionless flux formula
\beq
 \phi=\frac{BS}{\phi_0} \,.
\eeq
%Thus, \eqref{TRTRXi} reproduces the commutation rule \eqref{newTRTR}.
Therefore, the generalized commutation rule \eqref{TRTRXi} correctly reproduces the gauge-invariant expression given in \eqref{newTRTR}.

A general form for the composition rule can also be derived. As a generalization of \eqref{TRTRsym}, we have
\beq
\Tm{R_1}\Tm{R_2}=\Exp{ \frac{2\pi i}{\phi_0}\Lambda(\bm{R_1},\bm{R_2}) }\Tm{R_2+R_1}\,, \label{TRTRLMD}
\eeq
\begin{eqnarray}
\Lambda(\bm{R_1},\bm{R_2})&=&\xi(\bm{x},\bm{R_1})+\xi(\bm{x+R_1},\bm{R_2})-\xi(\bm{x},\bm{R_1+R_2}) \nn\\
&=& \xi(\bm{R_1},\bm{R_2})\,.
\end{eqnarray}
The commutation relation as a general form of \eqref{comsym} is then
\beq
[\Tm{R_1},\Tm{R_2}]=\left\{ \Exp{2\pi i\frac{\xi(\bm{R_1},\bm{R_2})}{\phi_0}} 
-\Exp{2\pi i\frac{\xi(\bm{R_2},\bm{R_1})}{\phi_0}}   \right\} \Tm{R_1+R_2}\,.
\eeq
Let us verify consistency with the specific gauge expressions discussed earlier. From \eqref{Lxi12}, we have
\begin{eqnarray}
\xi(\bm{R_1},\bm{R_2})&=&n_1m_2\xi(\bm{e_1},\bm{e_2}) +m_1n_2\xi(\bm{e_2},\bm{e_1})  \\
&=&\alpha B R_{1x}R_{2y} +(\alpha-1)BR_{1y}R_{2x} \,.
\end{eqnarray}
Thus, for the symmetric gauge where $\alpha=\frac{1}{2}$, we substitute
\beq
\xi(\bm{R_1},\bm{R_2})=\frac{B}{2}(\bm{R_1}\times\bm{R_2})\cdot\bm{n} =\frac{1}{2}BS
\eeq
to reproduce the composition rule \eqref{TRTRsym} and commutation relation \eqref{comsym} (note the reversed phase). For the Landau gauge, setting $\alpha=1$, we get
\beq
\xi(\bm{R_1},\bm{R_2})=n_1m_2\xi(\bm{e_1},\bm{e_2})=B R_{2y} R_{1x}
\eeq
which reproduces \eqref{TRTRldu} (note the reversed phase).

Looking at the geometric meaning of the composition rule \eqref{TRTRLMD}, similar to \eqref{4angle}, the composition rule can be interpreted as a circulation operation around a triangle formed by connecting the endpoints of $\bm{R}_1$ and $\bm{R}_1+\bm{R}_2$:
\beq
\Tm{R_1+R_2}^{-1}\Tm{R_1}^{}\Tm{R_2}^{}=e^{2\pi i \phi_3}\,,\quad  \phi_3
=\frac{\xi(\bm{R_1},\bm{R_2})}{\phi_0}\,.   \label{3angle}
\eeq
Also, decomposing the circulation around a parallelogram into two triangle circulation operations, \eqref{4angle} can be derived from \eqref{3angle}:
\begin{eqnarray}
e^{2\pi i \phi}&=&\Tm{R_1}^{-1}\Tm{R_2}^{-1}\Tm{R_1+R_2}\Tm{R_1+R_2}^{-1}\Tm{R_1}^{}\Tm{R_2}^{} \nn\\
&=&(\Tm{R_1+R_2}^{-1} \Tm{R_2}^{}\Tm{R_1}^{})^{-1}\Tm{R_1+R_2}^{-1}\Tm{R_1}^{}\Tm{R_2}^{} \nn \\
&=&e^{-2\pi i \phi_3(\bm{R_2},\bm{R_1})}e^{2\pi i \phi_3(\bm{R_1},\bm{R_2})}\,.
\end{eqnarray}
Thus, the relationship between $\phi$ and $\phi_3$ is, as can be seen from \eqref{XI} and \eqref{4angle}:
\beq
\phi=\phi_3(\bm{R_1},\bm{R_2})-\phi_3(\bm{R_2},\bm{R_1})\,.
\eeq

In this section, we derived the algebraic relations of MT operators without fixing the parameter $\alpha$, by keeping the gauge function $\xi$ in a parametric form. However, it is important to emphasize that this does not eliminate the $\xi$-dependence; rather, it simply refrains from specifying $\alpha$ or $\xi$ explicitly. In the next section, we introduce a gauge-invariant formulation of MT operators which, unlike the conventional definition, eliminates the explicit dependence on the gauge function $\xi$.

%%%%%%%%%%%%%%%%%%%%%%%%%%%%%%%%%%%%%%%%%%%%%%%%%%%%%%%%%%%%%%%
%     4.   Gauge invariant definition of MT
%%%%%%%%%%%%%%%%%%%%%%%%%%%%%%%%%%%%%%%%%%%%%%%%%%%%%%%%%%%%%%%%
\section{Gauge invariant MT }\label{sec:newTR}
\setcounter{equation}{0}
\indent

Thus far, we have identified two issues: first, the composition rule in the Landau gauge~\eqref{TRTRldu} does not coincide with that in the symmetric gauge~\eqref{TRTRsym}; second, as seen in the right-hand side of~\eqref{anyTmR}, the phase factor of $\Tm{R}$ varies depending on the gauge choice. For instance, in the Landau gauge, the expression~\eqref{LandTR} includes the gauge-dependent factor
\begin{equation}
\Exp{i\pi(2\alpha-1)\frac{B}{\phi_0}R_xR_y} \,, \label{Bfactor}
\end{equation}
while in the symmetric gauge~\eqref{TRsymbeta}, this factor reduces to unity. These discrepancies arise from the gauge-dependent nature of the conventional definition~\eqref{TmR}, which explicitly involves $\xi$ and $p_\mu$.

In Section~\ref{sec:TRdef}, we address this issue by constructing a gauge-invariant commutative group, thereby formulating a new definition of MT that is gauge-invariant. In Section~\ref{sec:genTR}, we further discuss its generalization.

%%%%%%%%%%%%%%%%%%%
%  4.1  definition
%%%%%%%%%%%%%%%%%%%
\subsection{Definition}\label{sec:TRdef}
\indent

Since the gauge-invariant form of $\Tm{R}$ is given by \eqref{TRsymbeta}, we consider expressing it in terms of $P_\mu$ defined in \eqref{covP}, instead of the gauge-dependent $p_\mu$. 
To this end, we focus on the following relations:
\begin{align}
&(\bm{R}\times\bm{\beta})_3=\frac{1}{2}(\bm{R}\times\bm{x})_3+\frac{l_B^2}{\hbar}\bm{R}\cdot\bm{P} \label{RbetaRx1} \\
&[ (\bm{R}\times\bm{x})_3, \bm{R}\cdot\bm{P}]=0\,,  \label{RbetaRx2}
\end{align}
from which it follows that
\beq
\Exp{\frac{i}{l_B^2}(\bm{R}\times\bm{\beta})_3}=\Exp{\frac{i}{2l_B^2}(\bm{R}\times\bm{x})_3}
\Exp{ \frac{i}{\hbar}\bm{R}\cdot\bm{P}} \,.
\eeq
Furthermore, using the identity
\beq
(\bm{R}\times\bm{x})_3=(\bm{R}\times\bm{x})\cdot\bm{n}=(\bm{n}\times\bm{R})\cdot\bm{x}
=\frac{1}{B}(\bm{B}\times\bm{R})\cdot\bm{x}     \label{Rx3}
\eeq
we obtain the following gauge-invariant expression:
\beq
\Exp{\frac{i}{l_B^2}(\bm{R}\times\bm{\beta})_3}=\Exp{\frac{i\pi}{\phi_0}(\bm{B}\times\bm{R})\cdot\bm{x}}
\Exp{ \frac{i}{\hbar}\bm{R}\cdot\bm{P}}\,.  \label{covTR}
\eeq
Comparing the RHS of \eqref{covTR} with the definition \eqref{TmR} of $\mathscr{T}_R$, we observe that the translation operator $T_{\bm{R}}$ is replaced by $\Exp{ \frac{i}{\hbar}\bm{R}\cdot\bm{P}}=:\mathscr{F}_{\bm{R}}$, and the phase factor is modified from $\Exp{2\pi i\xi/\phi_0}$ to $\Exp{\frac{i\pi}{\phi_0}(B\times R)\cdot x}$. 
The operator $\mathscr{F}_{\bm{R}}$ constitutes a magnetically extended commutative translation operator. It has the same algebraic structure as the ordinary translation operator, but with the momentum replaced by the covariant one $\bm{\pi}$, accompanied by the same phase factor as in \eqref{covTR}.
Specifically, $\mathscr{F}_{\bm{R}}$ satisfies the following relations:
\begin{align}
&\mathscr{F}_{\bm{R}}=\Exp{ \frac{i}{\hbar}\bm{R}\cdot\bm{P}}
=\Exp{\frac{i\pi}{\phi_0}(\bm{B}\times\bm{R})\cdot\bm{x}}\Exp{ \frac{i}{\hbar}\bm{R}\cdot\bm{\pi}}\,, \label{FR} \\
&\mathscr{F}_{\bm{R_1}}\mathscr{F}_{\bm{R_2}}=\mathscr{F}_{\bm{R_2}}\mathscr{F}_{\bm{R_1}}\,,\quad
\mathscr{F}_{\bm{R_1}}\mathscr{F}_{\bm{R_2}}=\mathscr{F}_{\bm{R_1}+\bm{R_2}}\,.
\end{align}

Furthermore, the phase factor multiplying $\mathscr{F}_{\bm{R}}$ on the RHS of \eqref{covTR} represents the volume of the parallelepiped spanned by the vectors $\bm{B}$, $\bm{R}$, and $\bm{x}$. Physically, this corresponds to the magnetic flux penetrating the parallelogram defined by $\bm{R} \times \bm{x}$, and thus quantifies the flux strength. This phase factor corresponds to the previously introduced $\xi$ in the definition \eqref{TmR}.

In contrast to the earlier formulation involving $\xi$, whose interpretation is obscured by the gauge ambiguity of the vector potential, the present formulation replaces it with the magnetic flux strength, a quantity with clear and direct physical meaning. This phase factor originates from the second term in the definition \eqref{covP}, which introduces a magnetic extension to the canonical momentum $p_\mu$. It reflects the phase accumulated when an electron moves across a magnetic lattice — corresponding to a hopping term in the tight-binding picture.

As a consequence of the above considerations, the LHS of \eqref{covTR} can be expressed in a manifestly gauge-invariant form (i.e., independent of the choice of gauge) by using the transformation formula \eqref{Rx3}. This expression acquires a clear physical meaning as the magnetic flux penetrating the parallelogram spanned by $\bm{R} \times \bm{\beta}$:
\beq
\frac{i}{l_B^2}(\bm{R}\times\bm{\beta})_3 = \frac{2\pi i}{\phi_0} (\bm{B}\times\bm{R})\cdot\bm{\beta}\,.
\eeq
Accordingly, we define the gauge-invariant MT operator $\mathcal{T}_{\bm{R}}$ using the RHS of \eqref{covTR}, yielding the following formula:
\beq
\mathcal{T}_{\bm{R}} := \Exp{\frac{i\pi}{\phi_0}(\bm{B}\times\bm{R})\cdot\bm{x}} 
\Exp{\frac{i}{\hbar}\bm{R}\cdot\bm{P}}\,.
\label{TRdef}
\eeq

To verify this explicitly, let us consider the symmetric gauge, where
\beq
P_\mu =p_\mu =-i\hbar\partial_\mu\,, \quad \xi(\bm{x}, \bm{R}) = \frac{1}{2} (\bm{B} \times \bm{R}) \cdot \bm{x}\,.
\eeq
Under this choice, it is evident that the expression \eqref{TRdef} exactly matches \eqref{TmR}. Importantly, although derived in the symmetric gauge, this expression remains valid in any gauge due to its gauge-invariant structure. As a direct consequence of the definition, and starting from \eqref{covTR}, we obtain the gauge-independent operator representation:
\beq
\mathcal{T}_{\bm{R}} = \Exp{\frac{i}{l_B^2}(\bm{R} \times \bm{\beta})_3}
= \Exp{\frac{2\pi i}{\phi_0} (\bm{B} \times \bm{R}) \cdot \bm{\beta}}\,.
\label{TRbybeta}
\eeq

To confirm this in the Landau gauge, we use $\xi = -B R_y x$ and substitute into \eqref{LandTR}. In doing so, we find that the phase factor in \eqref{Bfactor} cancels, yielding perfect agreement with \eqref{TRbybeta}:
\begin{align}
\mathcal{T}_{\bm{R}} &= \Exp{-i\pi \frac{B}{\phi_0} R_x R_y} \Exp{\frac{2\pi i}{\phi_0} \xi} T_{\bm{R}} \nn \\
&= \Exp{-i\pi \frac{B}{\phi_0} R_x R_y} \Tm{R}
= \Exp{\frac{i}{l_B^2}(\bm{R} \times \bm{\beta})_3}\,.
\end{align}

The composition and commutation rules for the magnetic translation operators follow immediately from the commutation relation of $\beta_i$ given in \eqref{beta_com}, and the identity $(\bm{R} \times \bm{\beta})_3 = R_i \beta_j \epsilon^{ij}$:
\beq
[(\bm{R}_1\times\bm{\beta})_3, (\bm{R}_2 \times \bm{\beta})_3] = i l_B^2 S\,, \quad S = (\bm{R}_1 \times \bm{R}_2)_3\,.
\eeq
This leads to the same algebraic structure as obtained in Section~\ref{sec:magTr} for the symmetric gauge, but now shown to be valid in any gauge:
\begin{align}
&\mathcal{T}_{\bm{R}_1} \mathcal{T}_{\bm{R}_2} = \Exp{-\frac{i\pi}{\phi_0} B S} \mathcal{T}_{\bm{R}_1 + \bm{R}_2}\,, \\
&\mathcal{T}_{\bm{R}_1} \mathcal{T}_{\bm{R}_2} = \Exp{-\frac{2\pi i}{\phi_0} B S} \mathcal{T}_{\bm{R}_2} \mathcal{T}_{\bm{R}_1}\,.
\end{align}

%%%%%%%%%%%%%%%%%%%%%%%%
%  4.2   Generalization 
%%%%%%%%%%%%%%%%%%%%%%%%
\subsection{Generalization}\label{sec:genTR}
\indent

In this section, we construct a generalized MT group by systematically analyzing the fundamental requirements for the MT operator $\mathcal{T}_{\bm{R}}$. From the properties discussed in the previous sections, several essential features can be identified:  
(i) the existence of a nontrivial commutative translation group;  
(ii) the emergence of a phase factor reflecting its magnetic extension;  
(iii) the formation of a gauge-invariant noncommutative translation group through composition as demonstrated in equation \eqref{RbetaRx1};  
and (iv) the necessity of condition \eqref{RbetaRx2} to enable the separation of the phase and operator parts.

These features can be reformulated in terms of the fundamental commutation relations among the elementary operators:
\begin{enumerate}[(i)]
\item \quad$ [P_i,P_j]=0\,, $\label{elem1} 
\item \quad$ \beta_i=\frac{1}{2}x_i-\frac{l_B^2}{\hbar}\eps^{ij}P_j\,,\quad
        [\beta_i,P_j]=\frac{i}{2}\hbar\delta_{ij}\,,\quad   [\beta_i,x_j]=il_B^2\eps^{ij}\,, $\label{elem2}
\item \quad$ [\beta_i,\beta_j]=il_B^2\eps^{ij}\,,$ \label{elem3} 
\item \quad$ [x_i,P_j]=i\hbar\delta_{ij}\,. $\label{elem4}
\end{enumerate}

It should be noted that the first equation in \eqref{elem2} is derived through the elimination of $\pi_i$ from \eqref{covP} using \eqref{defbeta}. Furthermore, multiplying this equation by $\eps^{ki}R_k$ and summing over $k$ yields \eqref{RbetaRx1}. Importantly, when the pair $(x_i,P_j)$ satisfies both \eqref{elem1} and \eqref{elem4}, then the second and third equations of \eqref{elem2} and the commutator \eqref{elem3} follow automatically as consequences of the first equation of \eqref{elem2}.

Our objective is to extend the original set of operators $U_0=\{x_i, P_i, \beta_i\}$ to a more general set $U=\{\chi_i,\tilde{P}_i,\tilde{B}_i\}$. In the general case, however, $U$ may be subject to constraints, necessitating additional flexibility in the coefficients. From a technical perspective, such flexibility becomes particularly important in situations where, for example, the commutation relation \eqref{elem3} is known a priori, while \eqref{elem1} and \eqref{elem4} have yet to be established. In such cases, adjustments to the coefficients in the relation \eqref{elem2} may be required in order to derive the commutation relations \eqref{elem1} and \eqref{elem4}. To accommodate such needs, we introduce a dimensionless arbitrary parameter $\varphi$ and propose the following generalized set of commutation relations:
\begin{eqnarray}
&& [\tilde{P}_i,\tilde{P}_j]=0\,, \label{elemg1} \\
&& \tilde{B}_i=\frac{1}{2}\varphi \chi_i-\frac{l_B^2}{\hbar}\eps^{ij}\tilde{P}_j\,,\quad
 [\tilde{B}_i,\tilde{P}_j]=\frac{i}{2}\varphi \hbar\delta_{ij}\,,\quad   [\tilde{B}_i,\chi_j]=il_B^2\eps^{ij}\,, \label{elemg2} \\
 &&[\tilde{B}_i,\tilde{B}_j]=i\alpha\eps^{ij}\,,\quad  \alpha=\varphi l_B^2\,,  \label{elemg3}  \\ 
 &&[\chi_i,\tilde{P}_j]=i\hbar\delta_{ij}\,. \label{elemg4}
\end{eqnarray}
It should be emphasized that $\tilde{B}$ possesses the same dimensional properties as $l_B$ (length dimension 1). Although we employ the same symbol $\alpha$ as the gauge parameter, this should not lead to confusion in the present context, as our discussion is entirely gauge-independent.
 
With these generalized relations, we extend \eqref{RbetaRx1} and \eqref{RbetaRx2} to more general forms:
\begin{align}
&(\bm{R}\times\tilde{\bm{B}})_3=\frac{1}{2}\varphi (\bm{R}\times\bm{\chi})_3+\frac{l_B^2}{\hbar}\bm{R}\cdot\tilde{\bm{P}}\,, \label{RbetaRX1} \\
&[ (\bm{R}\times\bm{\chi})_3, \bm{R}\cdot\tilde{\bm{P}}]=0\,.  \label{RbetaRX2}
\end{align}
When $\tilde{\bm{B}}$ satisfies the commutation relation \eqref{elemg3}, it follows that
\beq
[(\bm{R_1}\times\tilde{\bm{B}})_3,(\bm{R_2}\times\tilde{\bm{B}})_3]=i\alpha S\,,\quad 
 (\bm{R}\times\tilde{\bm{B}})_3=R_i\tilde{B}_j\eps^{ij}\,.
\eeq
Based on this structure, and in accordance with the forms \eqref{TRdef} and \eqref{TRbybeta}, we define the generalized magnetic translation (GMT) operator. Letting $L$ be a constant with the dimension of length squared, we write:
\begin{align}
\mathcal{T}_{\bm{R}}(\tilde{\bm{B}})
&:=\Exp{\varphi\frac{i\pi l_B^2}{L\phi_0}(\bm{B}\times\bm{R})\cdot\bm{\chi} }
\Exp{\frac{il_B^2}{L\hbar}\bm{R}\cdot\bm{\tilde{P}}} \\
&=\Exp{\frac{i}{L}(\bm{R}\times\tilde{\bm{B}})_3}=\Exp{\frac{i}{LB}(\bm{B}\times\bm{R})\cdot\tilde{\bm{B}} }\,.  \label{TRgen}
\end{align}
Here, the second line utilizes the identity \eqref{Rx3}. The operator $\mathcal{T}_{\bm{R}}(\tilde{\bm{B}})$ thus defined obeys the following composition and commutation relations:
\begin{align}
&\mathcal{T}_{\bm{R_1}}(\tilde{\bm{B}})\mathcal{T}_{\bm{R_2}}(\tilde{\bm{B}})=\Exp{-\frac{i\alpha}{2L^2} S}
\mathcal{T}_{\bm{R_1}+\bm{R_2}}(\tilde{\bm{B}}) \,,    \label{TSTS1} \\
&\mathcal{T}_{\bm{R_1}}(\tilde{\bm{B}})\mathcal{T}_{\bm{R_2}}(\tilde{B})=\Exp{-\frac{i\alpha}{L^2} S}
\mathcal{T}_{\bm{R_2}}(\tilde{\bm{B}})\mathcal{T}_{\bm{R_1}}(\tilde{\bm{B}}) \,.   \label{TSTS2}
\end{align}

For subsequent applications, we present several useful formulas. For distinct sets of operators $\bm{\Theta}$ and $\bm{\Theta}'$ with the commutation relation:
\beq
[\Theta_i, \Theta_j']=i\alpha \eps^{ij}\,.
\eeq
From this, we obtain the identity:
\beq
[(\bm{R_1}\times\bm{\Theta})_3,(\bm{R_2}\times\bm{\Theta}')_3]=i\alpha S\,,\quad 
 (\bm{R}\times\bm{\Theta}')_3=R_i\Theta_j'\eps^{ij} \,.
\eeq
Using the definition of the GMT operator \eqref{TRgen}, we find that the product of two such operators yields:
\begin{align}
\mathcal{T}_{\bm{R}_1}(\bm{\Theta})\, \mathcal{T}_{\bm{R}_2}(\bm{\Theta}')
&= \exp\left(-\frac{i\alpha}{2L^2} S\right)
\exp\left( \frac{i}{L} \left\{ (\bm{R}_1 \times \bm{\Theta})_3 + (\bm{R}_2 \times \bm{\Theta}')_3 \right\} \right)\,.
\end{align}
From this expression, the commutation and composition relations for the generalized operators follow straightforwardly:
\begin{align}
\mathcal{T}_{\bm{R}_1}(\bm{\Theta})\, \mathcal{T}_{\bm{R}_2}(\bm{\Theta}')
&= \exp\left( -\frac{i\alpha}{L^2} S \right)
\mathcal{T}_{\bm{R}_2}(\bm{\Theta}')\, \mathcal{T}_{\bm{R}_1}(\bm{\Theta})\,, \\
\mathcal{T}_{\bm{R}}(\bm{\Theta})\, \mathcal{T}_{\bm{R}}(\bm{\Theta}')
&= \mathcal{T}_{\bm{R}}(\bm{\Theta} + \bm{\Theta}')\,.
\end{align}

%%%%%%%%%%%%%%%%%%%%%%%%%%%%%%%%%%%%%%%%%%%%%%%%%%%%%%%%%%%%%%%%%%
%      5.  Examples
%%%%%%%%%%%%%%%%%%%%%%%%%%%%%%%%%%%%%%%%%%%%%%%%%%%%%%%%%%%%%%%%%%
\section{Examples}\label{sec:ex}
\setcounter{equation}{0}
\indent

In this section, we illustrate concrete realizations of the gauge-invariant generalized magnetic translation (GMT) operator introduced in equation~\eqref{TRgen}. When the dimensionless parameter is set to $\varphi=1$, and we choose $L=l_B^2$ along with the fundamental operator set $U_0 = \{x_i, P_i, \beta_i\}$, the construction naturally recovers the original magnetic translation operator $\mathcal{T}_{\bm{R}}(\tilde{\bm{B}})=\mathcal{T}_{\bm{R}}$. The terminology "magnetic translation" originates from the underlying translation group structure associated with charged particles in a magnetic field. 
In the case $\varphi\neq1$, no widely accepted naming convention exists. In this paper, we continue to refer to such operators as generalized magnetic translation (GMT) operators, with an emphasis on their gauge-invariant construction.

The following subsections provide examples of GMT operators with $\varphi\neq1$, illustrating the mathematical flexibility and physical implications of the general framework.

%%%%%%%%%%%%%%%%%%%%%%%%%%
%   5.1   varphi=0
%%%%%%%%%%%%%%%%%%%%%%%%%%
\subsection{The $\varphi=0$ case}\label{sec:phi0}
\indent

First, we consider the special case $\varphi = 0$ (i.e., $\alpha = 0$). In this regime, the GMT operator $\mathcal{T}_{\bm{R}}(\tilde{\bm{B}})$ defines a commutative group:
\begin{align}
&\mathcal{T}_{\bm{R}}(\tilde{\bm{B}})=\Exp{\frac{i}{l_B^2}(\bm{R}\times\tilde{\bm{B}})_3}=
\Exp{\frac{i}{\hbar} \bm{R}\cdot\tilde{\bm{P}}}\,,\label{TRB0} \\
&[(\bm{R_1}\times\tilde{\bm{B}})_3,(\bm{R_2}\times\tilde{\bm{B}})_3]=0\,,\quad 
\end{align}
where the operator set $\{\tilde{B}_i, \tilde{P}_i, \chi_i\}$ must satisfy the following commutation relations:
\begin{eqnarray}
&& [\tilde{P}_i,\tilde{P}_j]=0\,,\quad  [\tilde{B}_i,\tilde{B}_j]=0\,,\quad  [\tilde{B}_i,\tilde{P}_j]=0\,, \\
&& \tilde{B}_i=-\frac{l_B^2}{\hbar}\eps^{ij}\tilde{P}_j\,,\quad
    [\tilde{B}_i,\chi_j]=il_B^2\eps^{ij}\,,\quad [\chi_i,\tilde{P}_j]=i\hbar\delta_{ij}\,. 
\end{eqnarray}

One explicit realization of these relations employs the rotation operator $\mathcal{J}_3$ and the scale (dilation) operator $\mathcal{D}$ introduced in equation~\eqref{cylinDJ3}. Using them, we define $\tilde{B}_i^{(0)}$ and $\tilde{P}^{(0)}_i$ as:
\begin{align}
&\tilde{B}^{(0)}_1=i\frac{l_B}{\hbar}\mathcal{D}+il_B\tau^n\,,\quad \tilde{B}^{(0)}_2=-\frac{l_B}{\hbar}\mathcal{J}_3+l_B\theta^m\,, \\
&\tilde{P}^{(0)}_1=\frac{1}{l_B}(-\mathcal{J}_3+\hbar\theta^m)\,,\quad \tilde{P}^{(0)}_2=-\frac{i}{l_B}(\mathcal{D}+\hbar\tau^n)\,,\\
&(\chi_1,\chi_2)=(-l_B\theta, il_B\tau)\,,
\end{align}
where $n$ and $m$ are arbitrary integers. In the following, we take the representative choice $n = m = 1$ for concreteness.
Since $\mathcal{J}_3$ and $\mathcal{D}$ generate translations in the cylindrical coordinate space $(\theta, \tau)$, the GMT operator $\mathcal{T}_{\bm{R}}(\tilde{\bm{B}}^{(0)})$ defines a commutative translation group on this surface:
\beq
\mathcal{T}_{\bm{R}}(\tilde{\bm{B}}^{(0)})=\Exp{\frac{i}{\hbar}\bm{R}\cdot\tilde{\bm{P}}^{(0)}}\,.
\eeq
This commutative structure serves as a natural starting point for constructing noncommutative magnetic translations by generalizing to the case $\varphi \ne 0$. In such cases, $\tilde{B}^{(0)}_i$ is replaced by noncommutative translation generators $\tilde{B}_i$, and the resulting GMT operators form a nonabelian group structure on the same cylindrical geometry. Before turning to these noncommutative deformations, we explore further mathematical and physical aspects of the commutative subgroup realized at $\varphi = 0$.

Let us now introduce a new set of operators $\tilde{\bm{B}}^\mu$ and $\bm{\Theta}^\mu$ with $\mu = \tau, \theta, \pm$. These are linear combinations of the original translation generators $\tilde{B}_i^{(0)}$ and their canonical partners $\chi_i$. The motivation for this transformation is to decompose the GMT operator $\mathcal{T}_{\bm{R}}(\tilde{\bm{B}}^{(0)})$ into products of translation operators that obey the exchange and composition relations \eqref{TSTS1} and \eqref{TSTS2}, characteristic of gauge-invariant MTs. This decomposition is explicitly illustrated in equations \eqref{TRB0tauth} and \eqref{TRB0pm} below.

We first define the operators as follows:
\begin{align}
&\bm{\Theta}^\theta=(\Theta^\theta_1,\Theta^\theta_2) =(\frac{\hbar}{l_B}\theta, -\frac{1}{l_B}\mathcal{J}_3 )\,,\quad
 [\mathcal{J}_3,\theta]=-i\hbar  \label{th1} \\
&\bm{\Theta}^\tau=(\Theta^\tau_1,\Theta^\tau_2) =(-i\frac{\hbar}{l_B}\tau, -\frac{i}{l_B}\mathcal{D} )\,,\quad
 [\mathcal{D},\tau]=-i\hbar  \label{th2} \\
&\bm{\Theta}^\pm=\bm{\Theta}^\tau\pm\bm{\Theta}^\theta=(-i\frac{\hbar}{l_B}(\tau\pm i\theta), 
-\frac{2i}{l_B}\mathcal{D}_\pm) \,,\quad \mathcal{D}_\pm=\frac{1}{2}(\mathcal{D}\mp i\mathcal{J}_3)\,, \label{th3}
\end{align}
and relate them to the corresponding translation generators:
\beq
\tilde{B}_i^\mu=-\frac{l_B^2}{\hbar}\eps^{ij}\Theta^\mu_j\,,\quad 
\tilde{B}_i^\pm=\tilde{B}_i^\tau\pm\tilde{B}_i^\theta\,,\quad\quad(\mu=\tau,\theta) \,. \label{Theta2B}
\eeq
These operator sets obey the following commutation relations:
\begin{align}
&[\Theta_i^\mu,\Theta_j^\nu]=-i\frac{\hbar^2}{l_B^2}\eps^{ij}\delta_{\mu\nu}\,,\quad
[\tilde{B}_i^\mu,\tilde{B}_j^\nu]=-il_B^2\eps^{ij}\delta_{\mu\nu}\,,\quad
[\tilde{B}_i^\mu,\Theta_j^\nu]=-i\hbar\delta_{ij}\delta_{\mu\nu}\,,\label{TTmnij} \\
&[\Theta^\pm_i,\Theta^\pm_j]=-2i\frac{\hbar^2}{l_B^2}\eps^{ij}\,,\quad
[\tilde{B}_i^\pm,\tilde{B}_i^\pm]=-2il_B^2\eps^{ij}\,,\quad
[\tilde{B}_i^\pm,\Theta^\pm_j]=-2i\hbar\delta_{ij}\,,\\
&[\Theta^\pm_i,\Theta^\mp_j]=0\,,\quad
[\tilde{B}_i^\pm,\tilde{B}_i^\mp]=0 \,,\quad [\tilde{B}_i^\pm,\Theta^\mp_j]=0\,.
\label{TTpmij}
\end{align}

Based on these relations, the GMT operator can be recast in a compact exponential form. In doing so, it is convenient to define a new notation
\begin{align}
\hat{T}_{\bm{R}}[\bm{\Theta}^\mu] := \exp\left( \frac{i}{\hbar} \bm{R} \cdot \bm{\Theta}^\mu \right), \label{trbmu1}
\end{align}
where the bracket $[\bm{\Theta}^\mu]$ explicitly indicates that the generator of translation depends functionally on the operator vector $\bm{\Theta}^\mu$. It is to emphasize that the translation is generated via $\bm{\Theta}^\mu$ through inner product with $\bm{R}$. With this convention, the operator $\mathcal{T}_{\bm{R}}(\tilde{\bm{B}}^\mu)$ takes the form
\begin{align}
\mathcal{T}_{\bm{R}}(\tilde{\bm{B}}^\mu) 
= \exp\left( \frac{i}{l_B^2} (\bm{R} \times \tilde{\bm{B}}^\mu)_3 \right) 
= \hat{T}_{\bm{R}}[\bm{\Theta}^\mu]\,. 
\end{align}
In particular, for $\mu = \pm$, we find that
\begin{align}
\mathcal{T}_{\bm{R}}(\tilde{\bm{B}}^\pm) 
= \hat{T}_{\bm{R}}[\bm{\Theta}^\pm] 
= \hat{T}_{\bm{R}}[\bm{\Theta}^\tau] \, \hat{T}_{\pm \bm{R}}[\bm{\Theta}^\theta]\,. \label{trbmu2}
\end{align}

Using this notation, from \eqref{TRB0} we get the decomposition of $\mathcal{T}_{\bm{R}}(\tilde{\bm{B}}^{(0)})$ in several equivalent forms:
\begin{align}
\mathcal{T}_{\bm{R}}(\tilde{\bm{B}}^{(0)}) 
&= \hat{T}_{(R_1, R_1)}[\bm{\Theta}^\tau] \, \hat{T}_{(R_2, R_2)}[\bm{\Theta}^\theta], \label{TRB0tauth} \\
&= \hat{T}_{(R_1, R_1)}[\bm{\Theta}^+] \, \hat{T}_{(R_2 - R_1, R_2 - R_1)}[\bm{\Theta}^\theta], \\
&= \hat{T}_{(R_1, R_1)}[\bm{\Theta}^-] \, \hat{T}_{(R_1 + R_2, R_1 + R_2)}[\bm{\Theta}^\theta].
\end{align}
These expressions reveal that the commutative GMT group action can be decomposed into two consecutive translations along 45-degree spiral directions in the $(\theta,\tau)$-plane. Notably, for $R_2 = \pm R_1$, from \eqref{trbmu2} and \eqref{TRB0tauth} we can write:
\begin{align}
\mathcal{T}_{(R_1, R_2)}(\tilde{\bm{B}}^{(0)}) 
= \hat{T}_{(R_1, R_1)}[\bm{\Theta}^\pm], \label{TRB0pm}
\end{align}
in which case the GMT operator is generated directly by the operators $\mathcal{D}_\pm$, as $\Theta_2^\pm = -\frac{2i}{l_B} \mathcal{D}_\pm$ per equation~\eqref{th3}. This highlights the explicit role of the spiral dilations $\mathcal{D}_\pm$ in characterizing certain symmetry directions of the commutative GMT group when the commutative translation $\tilde{\bm{B}}^{(0)}$ takes special directions.

While the investigations so far are merely decompositions into commutative components, the operator $\hat{T}_{\bm{R}}[\bm{\Theta}^\mu]$ turns out to exhibit interesting algebraic properties. Although $\hat{T}_{\bm{R}}[\bm{\Theta}^\mu]$ satisfies only part of the relations \eqref{elemg1}–\eqref{elemg4}, it retains the essential group structure of MT. Specifically, setting $\alpha=-l_B^2$ in the RHS of \eqref{elemg3} and taking $L=l_B^2$, the composition and exchange rules from \eqref{TSTS1} and \eqref{TSTS2} become (note Eq.\eqref{LB}):
\begin{align}
\hat{T}_{\bm{R_1}}[\bm{\Theta}^\mu]\hat{T}_{\bm{R_2}}[\bm{\Theta}^\mu]
&= \exp\left(\frac{i\pi}{\phi_0} BS\right)\hat{T}_{\bm{R_1}+\bm{R_2}}[\bm{\Theta}^\mu]\,, \label{TJTJ1} \\
\hat{T}_{\bm{R_1}}[\bm{\Theta}^\mu]\hat{T}_{\bm{R_2}}[\bm{\Theta}^\mu]
&= \exp\left(\frac{2\pi i}{\phi_0} BS\right)\hat{T}_{\bm{R_2}}[\bm{\Theta}^\mu]\hat{T}_{\bm{R_1}}[\bm{\Theta}^\mu]\,. \label{TJTJ2}
\end{align}

Moreover, for the combinations $\bm{\Theta}^\pm = \bm{\Theta}^\tau \pm \bm{\Theta}^\theta$, where the structure constant effectively doubles, the corresponding operators satisfy:
\begin{align}
&\hat{T}_{\bm{R_1}}[\bm{\Theta}^\pm]\hat{T}_{\bm{R_2}}[\bm{\Theta}^\pm]=\Exp{\frac{2\pi i}{\phi_0} BS}
\hat{T}_{\bm{R_1}+\bm{R_2}}[\bm{\Theta}^\pm] \,,    \label{TJTJ1pm} \\
&\hat{T}_{\bm{R_1}}[\bm{\Theta}^\pm]\hat{T}_{\bm{R_2}}[\bm{\Theta}^\pm]=\Exp{\frac{4\pi i}{\phi_0} BS}
\hat{T}_{\bm{R_2}}[\bm{\Theta}^\pm]\hat{T}_{\bm{R_1}}[\bm{\Theta}^\pm] \,.   \label{TJTJ2pm}
\end{align}
In light of these algebraic structures, it is natural to regard $\hat{T}_{\bm{R}}[\bm{\Theta}^\mu]$ for $\mu = \tau, \theta, \pm$ as a generalized GMT in a broad sense, extending the conventional concept to operators generated by $\bm{\Theta}^\mu$, not just by position or momentum.

%%%%%%%%%%%%%%%%%%%%%%%%%%%%%%
%   5.2   varphi=2
%%%%%%%%%%%%%%%%%%%%%%%%%%%%%%
\subsection{The $\varphi=2$ case}\label{sec:newTR2}
\indent

Building on our analysis of the commutative case ($\varphi = 0$), we have shown that the translation operators $\mathcal{T}_{\bm{R}}(\tilde{\bm{B}}^{(0)})$ can be described in terms of the operator group $\hat{T}_{\bm{R}}[\bm{\Theta}^\mu]$, which shares the same algebraic structure as the MT operators. Regarding the definition of GMT, if we demand all the physical properties \eqref{elem1}–\eqref{elem4} of MT, then the full set of conditions \eqref{elemg1}–\eqref{elemg4} must be satisfied (which we refer to as strict GMT). However, if we relax this requirement and focus solely on reproducing the algebraic (group) structure, then condition \eqref{elemg3} alone suffices (referred to as broad GMT).

In the following, we show that even for the $\varphi \ne 0$ case, the physical operators (strict GMT) can be constructed from these mathematically defined operators (broad GMT) as fundamental building blocks.

A solution to the strict GMT conditions \eqref{elemg1}-\eqref{elemg4} for $\varphi=2$ is:
\beq
\tilde{B}_i=\chi_i-\frac{l_B^2}{\hbar}\eps^{ij}\tilde{P}_j\,,\quad 
(\tilde{P}_1,\tilde{P}_2)=(-\frac{1}{l_B}\mathcal{J}_3,-\frac{i}{l_B}\mathcal{D}) \label{BPphi2}
\eeq
which satisfies \eqref{RbetaRX1}. Accordingly, the explicit form of the strict GMT operator is obtained by substituting \eqref{BPphi2} into:
\begin{align}
\mathcal{T}_{\bm{R}}(\tilde{\bm{B}})&=\Exp{\frac{i}{l_B^2}(\bm{R}\times\tilde{\bm{B}})_3}
=\Exp{\frac{2\pi i}{\phi_0}(\bm{B}\times\bm{R})\cdot\bm{\chi} }\Exp{\frac{i}{\hbar}\bm{R}\cdot\tilde{\bm{P}}} \\
&=\hat{T}_{(-R_1,R_2)}[\bm{\Theta}^\tau]\hat{T}_{(R_2,R_1)}[\bm{\Theta}^\theta]\,. \label{TRBisTHTH}
\end{align}
By setting $\alpha=2l_B^2$ and $L=l_B^2$ in \eqref{TSTS1} and \eqref{TSTS2}, the corresponding composition and exchange rules are given by:
\begin{align}
&\mathcal{T}_{\bm{R_1}}(\tilde{\bm{B}})\mathcal{T}_{\bm{R_2}}(\tilde{\bm{B}})=\Exp{-\frac{2\pi i}{\phi_0} BS}
\mathcal{T}_{\bm{R_1}+\bm{R_2}}(\tilde{\bm{B}}) \,,    \label{TJTJ1phi2} \\
&\mathcal{T}_{\bm{R_1}}(\tilde{\bm{B}})\mathcal{T}_{\bm{R_2}}(\tilde{\bm{B}})=\Exp{-\frac{4\pi i}{\phi_0} BS}
\mathcal{T}_{\bm{R_2}}(\tilde{\bm{B}})\mathcal{T}_{\bm{R_1}}(\tilde{\bm{B}}) \,.  \label{TJTJ2phi2}
\end{align}
These relations can also be directly verified using \eqref{TJTJ2} and \eqref{TRBisTHTH}.

Similar to the $\varphi=0$ case, to make the relationship with $\mathcal{D}_\pm$ more transparent, we switch to a coordinate representation in which the two components of $\tilde{B}_i$, $\tilde{P}_i$, and $\chi_i$ are reorganized into $\tilde{B}_\pm$, $\tilde{P}_\pm$, and $\chi_\pm$:
\beq
\tilde{B}_\pm=\frac{1}{\sqrt{2}}(\tilde{B}_2\pm\tilde{B}_1)=\chi_\pm \mp\frac{l_B^2}{\hbar}\tilde{P}_\mp\,,
\eeq
\beq
\chi_\pm=\frac{1}{\sqrt{2}}(\chi_2\pm\chi_1)=\frac{il_B}{\sqrt{2}}(\tau\pm i\theta)\,,\quad
\tilde{P}_\pm=\frac{1}{\sqrt{2}}(\tilde{P}_2\pm\tilde{P}_1)=-\frac{i}{l_B}\sqrt{2}\mathcal{D}_\pm\,.
\eeq
With this representation, \eqref{TRBisTHTH} can be rewritten in terms of $\Theta^\pm_i$:
\beq
\mathcal{T}_{\bm{R}}(\tilde{\bm{B}})=\hat{T}_{(-r_+,r_-)}[\bm{\Theta}^-]\hat{T}_{(r_-,r_+)}[\bm{\Theta}^+]\,, \label{TbyTpm}
\eeq
where
\beq
r_\pm=\frac{1}{\sqrt{2}}R_\pm=\frac{1}{2}(R_2\pm R_1)\,.
\eeq
A notable feature of \eqref{TRBisTHTH} or \eqref{TbyTpm} is that, due to \eqref{TTmnij} and \eqref{TTpmij}, the operator for $\varphi=2$ can be factorized into a product of two $\varphi=0$ type operators, each corresponding to a mutually commuting direction. These directions are associated with the degrees of freedom of rotation and dilatation, respectively.

In summary, when $\tilde{P}_i$ is taken to be the generator $P_i$ of the conformal group (with $\varphi=1$, $L=l_B^2$, and $\tilde{B}_i=\beta_i$), the MT operator is precisely given by $\mathcal{T}_{\bm{R}}$. In contrast, when $\tilde{P}_i$ corresponds to the generators of the similarity group $\mathfrak{sim}(2)$, namely $\mathcal{J}_3$, $\mathcal{D}$, or $\mathcal{D}_\pm$ (with $\varphi=2$, $L=l_B^2$, and $\tilde{B}_i$ defined as in \eqref{BPphi2}), the GMT operator is represented by $\mathcal{T}_{\bm{R}}(\tilde{\bm{B}})$. It is also conceivable to construct GMT operators using gauge-invariant realizations of other conformal generators, such as $\mathcal{K}_\pm$.

%%%%%%%%%%%%%%%%%%%%%%%%%%%%%%%%%%%%%%%%%%%%%%%%%%%%%%%%%%%
%     6.  Curtright-Zachos generator
%%%%%%%%%%%%%%%%%%%%%%%%%%%%%%%%%%%%%%%%%%%%%%%%%%%%%%%%%%%
\section{Curtright-Zachos generators}\label{sec:cz}
\setcounter{equation}{0}

In this section, we demonstrate that the differential operator representation of the CZ algebra can be expressed using the GMT defined in Section~\ref{sec:ex}. Based on these results, we discuss how the ($\mathcal{J}_3$,$\mathcal{D}$) duality explored in Section~\ref{sec:confg} is related to the internal symmetry of the FFZ algebra~\cite{FFZ}.

The differential operator representation of the CZ generator is given by the following linear combination of non-local differential operators~\cite{AS4}-\cite{AS2}:
\beq
\Op{t}{m}{k} = z^m q^{-k(z\partial + \frac{m}{2} + \Delta)}\,,  \label{That} 
\eeq
which satisfies operational relations equivalent to those of the MT operator. That is, if we write $\Op{t}{n}{k}=t_{(n,k)}$, it satisfies the exchange relation:
\beq
t_{(n,k)}t_{(m,l)}= q^{nl-mk}  t_{(m,l)}t_{(n,k)}\,, \label{trans} 
\eeq
and the composition relation:
\beq
t_{(n,k)}t_{(m,l)}= q^{\frac{nl-mk}{2}} t_{(n+m,k+l)}\,. \label{Trans}
\eeq
The operator $t_{(n,k)}$ satisfies the following commutation relation (FFZ algebra):
\beq
[t_{(n,k)}, t_{(m,l)}]= (q^{\frac{nl-mk}{2}}-q^{\frac{mk-nl}{2}} )t_{(n+m,k+l)}\,,
\eeq
which admits an internal symmetry (automorphism) under the transformation $t_{(n,k)}\lra t_{(k,n)}$ and $q\lra q^{-1}$ leaving the algebraic structure invariant, though this internal structure is not apparent from the form of \eqref{That}.

To unveil this mystery, we need to understand the internal structure of the MT operator. Although the exchange relation \eqref{TRTRXi} and composition relation \eqref{TRTRLMD} of the MT operator $\Tm{R}$ are similar to \eqref{trans} and \eqref{Trans}, to accurately discern their agreement, we take the symmetric gauge $\xi(\bm{e_1},\bm{e_2})=\frac{B}{2}a^2$ and set:
\beq
\Lambda=(n_1m_2-m_1n_2)\xi(\bm{e_1},\bm{e_2})\,,\quad \Xi=2(n_1m_2-m_1n_2)\xi(\bm{e_1},\bm{e_2})
\eeq
\beq
\bm{R_1}=(n_1,m_1)=(n,k)\,,\quad \bm{R_2}=(n_2,m_2)=(m,l)
\eeq
\beq
q=\Exp{2\pi i \frac{B a^2}{\phi_0}}=\Exp{i a^2 l_B^{-2}}
\eeq
With these identifications, the exchange relation \eqref{TRTRXi} and composition relation \eqref{TRTRLMD} become:
\begin{align}
&\Tm{R_1}\Tm{R_2}=q^{nl-mk}\Tm{R_2}\Tm{R_1}\,,  \label{exchangeTR}  \\
&\Tm{R_1}\Tm{R_2}=q^{\frac{nl-mk}{2}}\Tm{R_1+R_2}\,. \label{combineTR}
\end{align}
We recognize that these clearly match the forms of \eqref{trans} and \eqref{Trans}.

Thus, it appears reasonable to consider the non-local differential operator $\Op{t}{n}{k}$ as a realization of the MT operator $\Tm{R}$ (in the symmetric gauge). However, one question remains: the differential operation in $\Tm{R}$ is the translation operation $\hat{p}\sim\partial_z$, whereas the differential operation in $\Op{t}{n}{k}$ is $z\partial_z$, which is not a translation. This cannot be explained by simple magnetic translation on a plane. Furthermore, it is not particularly desirable for a symmetry like the CZ algebra to depend on gauge choice.

These concerns can be resolved by considering the gauge-invariant MT discussed in Section~\ref{sec:ex}. Noting the Campbell-Baker-Hausdorff formula:
\beq
e^Ae^B=e^{A+B+\frac{1}{2}[A,B]+\cdots}
\eeq
we have
\begin{eqnarray}
\Op{t}{n}{k} &=&
z^n q^{-k(z\partial +\frac{n}{2}+\Delta)} =e^{n\ln{z}}e^{-ia^2l_B^{-2}k(z\partial +\frac{n}{2}+\Delta)} \nn\\
&=&\Exp{\frac{i}{\hbar}\bm{\lambda}\cdot\bm{\Theta}}  \label{Tmtheta}
\end{eqnarray}
\beq
\bm{\lambda}=(nl_B,k\frac{a^2}{l_B})\,,\quad\quad 
\bm{\Theta}=(\Theta_1,\Theta_2)=(-i\frac{\hbar}{l_B}\ln{z}, -\frac{\hbar}{l_B}(z\partial+\Delta))\,.
\eeq

Let us consider the transformation of $z$ to the cylindrical surface $(\tau,\theta)$ as noted in \eqref{cylinDJ3}. The transformations to the unit circle $z=e^{i\theta}$, to the positive axis $z=e^\tau$, and to the cylindrical surface (including the negative axis) $z=e^{\tau\pm i\theta}$ yield respectively:
\begin{align}
&\bm{\Theta}=(\frac{\hbar}{l_B}\theta, -\frac{1}{l_B}\mathcal{J}_3 -\frac{\hbar}{l_B}\Delta)\,,  \label{Jdelta}\\
&\bm{\Theta}=(-i\frac{\hbar}{l_B}\tau, -\frac{i}{l_B}\mathcal{D}-\frac{\hbar}{l_B}\Delta )\,,
\label{Ddelta} \\
&\bm{\Theta}=(-i\frac{\hbar}{l_B}(\tau\pm i\theta), -\frac{2i}{l_B}\mathcal{D}_\pm-\frac{\hbar}{l_B}\Delta) \,. \label{JDdelta}
\end{align}

For $\Delta=0$, these correspond to $\bm{\Theta}^\theta$, $\bm{\Theta}^\tau$, and $\bm{\Theta}^\pm$ from equations \eqref{th1}-\eqref{th3}. Therefore, $\Op{t}{n}{k}$ \eqref{Tmtheta} is a broad GMT with $\varphi=0$ (or one direction separated from $\varphi=2$) by \eqref{trbmu1} and \eqref{trbmu2}, and we have:
\beq
\Op{t}{n}{k}=\mathcal{T}_{\bm{\lambda}}(\tilde{\bm{B}}^\mu)=\hat{T}_{\bm{\lambda}}[\bm{\Theta}^\mu]\,,
\quad\quad (\mu=\theta,\tau,\pm)\,,
\eeq
where $\bm{\tilde{B}}^\mu$ is given by $\bm{\Theta}^\mu$ through the transformation \eqref{Theta2B}.

Equation \eqref{Jdelta} corresponds to replacing $\mathcal{J}_3$ in $\bm{\Theta}^\theta$ with the sum of orbital and spin angular momenta:
\beq
\mathcal{J}_3=-i\hbar \partial_\theta = -i\hbar(x\partial_y-y\partial_x)
\,\ra\,  \mathcal{J}_3+\hbar\Delta\,.
\label{J3gen}
\eeq
The appearance of $\Delta$ in \eqref{Ddelta} and \eqref{JDdelta} likely reflects the ($\mathcal{J}_3$,$\mathcal{D}$) duality discussed in Section~\ref{sec:confg}. In this paper, since we do not consider spin, we may set $\Delta=0$.

Finally, let us comment on the duality of $(n,k)$. The parameter $n$ corresponds to coordinates on the cylindrical surface with translation unit $l_B$, and $k$ corresponds to angular momentum with unit $a^2l_B^{-1}$. Considering the $\mathcal{D}\lra\mathcal{J}_3$ symmetry, $n$ can be viewed as coordinates along a line on the $w$ or $\bar{w}$ plane ($w=\tau+ i\theta$) with translation unit $l_B$, while $k$ can be seen as angular momentum as an eigenvalue of $\mathcal{D}_\pm$. Since $n$ corresponds to an eigenvalue of $\mathcal{J}_3$, we can impose a periodic condition, and for phase angle $\frac{2\pi p}{N}$, we have cyclic representation~\cite{NQ} corresponding to $n=p+mN$, $p=0,1,2,..N-1$.

Putting everything together, we have shown that the non-local differential operator $\Op{t}{n}{k}$ can be interpreted as a GMT operator in a generalized sense, with its duality symmetry $(n, k) \leftrightarrow (k, n)$ and $q \leftrightarrow q^{-1}$ understood as a manifestation of the $\mathcal{J}_3 \leftrightarrow \mathcal{D}$ symmetry on a cylindrical geometry. This perspective not only clarifies the internal structure of the FFZ algebra but also reveals a deeper geometric origin of the CZ generators in terms of gauge-invariant magnetic translations.

%%%%%%%%%%%%%%%%%%%%%%%%%%%%%%%%%%%%%%%%%%%%%%%%%%%%%%%%%%%
%     7.  Conclusions & Discussions
%%%%%%%%%%%%%%%%%%%%%%%%%%%%%%%%%%%%%%%%%%%%%%%%%%%%%%%%%%%
\section{Conclusions and Discussions}\label{sec:end}
\setcounter{equation}{0}

In this paper, we have proposed a gauge-invariant formulation of magnetic translation (MT) operators and Curtright-Zachos (CZ) generators in a two-dimensional electron system under a uniform magnetic field (Landau problem). Conventional MT operators are known to be gauge-dependent, which causes their algebraic relations to vary under gauge transformations. To overcome this issue, we introduced a gauge-invariant operator -- the generalized magnetic translation (GMT) -- and systematically studied its algebraic and physical properties.

Our main contributions are summarized in three key points:

1. Gauge-invariant definition of magnetic translation operators (GMT): 
We constructed GMT operators by replacing the canonical momentum $p_\mu$ with the gauge-invariant (pseudo) momentum $P_\mu$, and incorporating a gauge-invariant phase factor closely related to the Aharonov-Bohm phase. This construction eliminates gauge ambiguity and ensures a clear physical interpretation.

2. Gauge-invariant realization of CZ generators using GMT and $\mathfrak{so}(3,1)$ algebra: 
By combining the generators of a gauge-invariant $\mathfrak{so}(3,1)$ algebra — comprising pseudo dilatation, pseudo orbital angular momentum (OAM), pseudo conformal, and translation generators — with the GMT, we constructed gauge-invariant representations of CZ generators. The duality between the pseudo-dilatation and pseudo-OAM operators, embedded in the $\mathfrak{sim}(2)$ subalgebra, plays a crucial role in this construction.

3. Rigorous mathematical formulation: We formulated the MT and CZ operators with full mathematical rigor. By detailing the reverting transformation (Appendix~\ref{sec:Landau}) and gauge invariance (Appendix~\ref{sec:gauge}), we ensured that our framework remains consistent and broadly applicable.

Several avenues for future work emerge from our formulation. For example, when a pair of gauge-invariant commuting operators $(\pi, \beta)$ exists and commutes with the Hamiltonian $H$, the group element $\mathscr{F}_{\bm{R}}$ defined in \eqref{FR} becomes proportional to $\exp{\frac{i}{\hbar}\bm{R}\cdot\bm{\pi}}$, while the GMT operator $\mathcal{T}_{\bm{R}}$ from \eqref{covTR} takes the form $\exp{\frac{i}{l_B^2}(\bm{R}\times\bm{\beta})_3}$. A similar correspondence holds for the pair $(\mathcal{J}_3, \mathscr{D}_3)$, suggesting a general framework for defining GMTs. Beyond the $\mathfrak{sim}(2)$ subalgebra, gauge-invariant pairs like $(\mathcal{K}_+, \mathcal{K}_-)$ also exist. In such cases, their non-commutativity with $H$ introduces additional challenges, such as determining whether a suitable operator set $\{\chi_i, \tilde{P}_i, \tilde{B}_i\}$ exists, whether time-dependent symmetries are required, and whether the physical constraints need reformulation.

Regarding further elucidation of the mathematical structure of GMT: It is also important to investigate the algebra generated by GMT and the role of the CZ algebra from a mathematical perspective. In particular, studying the representation theory of GMT and its relationship with other infinite-dimensional algebras~\cite{JS,KS} presents an intriguing challenge.

From a physical perspective, GMTs offer an alternative framework for describing electron motion under magnetic fields. Their role in systems beyond the Landau problem — such as graphene and topological insulators~\cite{2dTIs} — deserves further investigation. In particular, clarifying their connection with time-dependent electric fields and extended symmetries~\cite{NH,ET} is an important direction.

The $W_\infty$ algebra and FFZ algebra, which have the symplectic form on $\mathbb{Z}^2$ lattice similar to the CZ algebra, play important roles in various aspects, such as the description of edge states in quantum Hall effects, the construction of operators in composite particle pictures, and the description of holography in gravity theory~\cite{ahn}. The gauge-invariant formulation obtained in this study is expected not only to deepen our understanding of the structure and physical interpretation of the CZ algebra but also to provide new perspectives and precise theoretical descriptions in related fields of research, serve as a useful tool in model construction based on quantum space and non-commutative geometry, and contribute to the development of new mathematical advances.

In summary, this study introduces a new gauge-invariant approach to MT operators and the CZ algebra, shedding light on their physical and mathematical structure. We hope that our results will stimulate further research across physics and mathematics.

%%%%%%%%%%%%%%%%%%%%%%%%%%%%%%%%%
%  CRediT statement
%%%%%%%%%%%%%%%%%%%%%%%%%%%%%%%%%
\section*{Declaration of generative AI and AI-assisted technologies in the writing process}
During the preparation of this work the author used Claude 3.7 Sonnet in order to improve grammar and enhance language expression. 
After using the service, the author reviewed and edited the content as needed and takes full responsibility for the content of the publication.

\section*{CRediT authorship contribution statement} 
 \textbf{Haru-Tada Sato:} Writing-Original Draft, Conceptualization, Methodology, Investigation, Validation.

\section*{Declaration of competing interest}
The author declares that we have no known competing financial interests or personal relationships that could have appeared to influence the work reported in this paper.

%\section*{Acknowledgements}
%The author would like to thank Professor N. Aizawa for the valuable discussions and %comments on this paper.

\section*{Data availability}
No data was used for the research described in the article. 

%%%%%%%%%%%%%%%%%%%%%%%%%%%%%%%%%%%%%%%%%%%%%%%%%%%%%%%%%%%%%%%%%%%
%                           A. 　2-dim electron QM 
\appendix
%%%%%%%%%%%%%%%%%%%%%%%%%%%%%%%%%%%%%%%%%%%%%%%%%%%%%%%%%%%%%%%%%%%
\setcounter{equation}{0}
\section{Reverting to arbitrary gauge}\label{sec:Landau}
\indent

When the symmetric gauge is chosen, the effects of the gauge potential may appear to vanish, which can improve the transparency of expressions. However, this often leads to confusion, as the resulting forms coincide with those of free electron systems. In this appendix, we first summarize important considerations regarding the distinctive nature of the symmetric gauge, and then present a method for deriving gauge-invariant expressions of physical operators by starting from this gauge choice.

%%%%%%%%%%%%%%%%%%%%%%%%%%%%%%
%   A.1    setup (notations)
%%%%%%%%%%%%%%%%%%%%%%%%%%%%%%
\subsection{The setup of the Landau problem}
\indent
We consider the Hamiltonian of a single-electron system confined to the $xy$ plane, subject to a magnetic field $\bm{B}=(0,0,B)$ perpendicular to the plane. We denote the cyclotron frequency by $\omega$, the magnetic length by $l_B$, the cyclotron center (aka the guiding center) by $\bm{\beta}$, and the magnetic flux quantum by $\phi_0$.
\beq
H=\frac{1}{2m} ({\bm p} +\frac{e}{c}\bm{A})^2 = \frac{1}{2m}\bm{\pi}^2 
\eeq
\beq
p_i= -i\hbar\partial_i\,,\quad
\pi_i = -i\hbar\partial_i  + \frac{e}{c}A_i\,
\eeq
\beq
\bm{B}=\nabla\times\bm{A}\,, \quad B_i=\eps^{ijk}\partial_j A_k\,,
\eeq
\beq
[\pi_i,\pi_j] = -i \frac{\hbar^2}{l_B^2}\eps_{ij}\,, \quad[x_i, \pi_j] = i\hbar \delta_{ij}\,,\quad  
l_B=\sqrt{\frac{\hbar c}{eB}} =\sqrt{\frac{\phi_0}{2\pi B}}  \label{LB}
\eeq
\beq
[a,a^\dagger]=1\,,\quad\ 
a=\frac{l_B}{\sqrt{2}\hbar} (\pi_2 + i \pi_1)\,,\quad  a^\dagger=\frac{l_B}{\sqrt{2}\hbar} (\pi_2 - i \pi_1)\,.
\eeq
\beq
\pi_1=\frac{-i\hbar}{\sqrt{2}l_B}(a-a^\dagger)\,,\quad \pi_2=\frac{\hbar}{\sqrt{2}l_B}(a+a^\dagger)\,.
\eeq
\beq
H=\hbar\omega(a^\dagger a+\frac{1}{2})\,, \quad \omega=\frac{eB}{mc}\,.  
\eeq
\beq
\bm{\beta} = \bm{x} - \frac{l_B^2}{\hbar}(\pi_2, -\pi_1)\,,\quad\mbox{or}\quad \beta_i=x_i-\frac{l_B^2}{\hbar}\eps^{ij}\pi_j
\label{defbeta}
\eeq
\beq
 [\beta_i, \beta_j] = il_B^2\eps^{ij}\,,\quad
[\beta_i, \pi_j] =0\,,\quad [\beta_i,x_j]=il_B^2\eps^{ij}\,.    \label{beta_com}
\eeq
\beq
\ra  [\bm{\beta}, H] = [\bm{\beta}, a] = [\bm{\beta}, a^\dagger]=0\,.
\eeq
This implies Landau degeneracy: different eigenvalues of $\beta$ for the same energy eigenvalue. $\bm{\beta}$ is a harmonic oscillator commuting with $a$, and can be expressed as follows, revealing that the eigenvalues of $\beta^2$ are quantized in units of $2l_B^2$:
\beq
[b,b^\dagger]=1\,,\quad\
b=\frac{1}{\sqrt{2}l_B} (\beta_1 + i \beta_2)\,,\quad  b^\dagger=\frac{1}{\sqrt{2}l_B} (\beta_1 - i \beta_2)\,,
\eeq
\beq
\beta^2 = 2l_B^2(b^\dagger b +\frac{1}{2})\,.
\eeq

Noting that the differential operator $\partial^\dagger=-\partial$, we have
\beq
\bm{p}^\dagger=\bm{p}\,,\quad
\bm{\pi}^\dagger=\bm{\pi}\,,\quad \bm{\beta}^\dagger=\bm{\beta}\,.
\eeq
The charge conjugate of the covariant momentum, $\bm{\pi}^c = \bm{p}-\frac{e}{c}\bm{A}$, satisfies
\beq
[\pi^c_i,\pi^c_j] = i \frac{\hbar^2}{l_B^2}\eps_{ij}\,,
\eeq
and its commutation relations with $\bm{\pi}$ and $\bm{x}$ are
\beq
 [\pi_i, \pi^c_j]=0\,,\quad \quad[x_i, \pi^c_j] = i\hbar \delta_{ij}\,.
\eeq

It is worth noting a particular feature of the symmetric gauge. When we choose the symmetric gauge:
\beq
A_i=-\frac{1}{2}B\eps^{ij}x_j \,,
\eeq
$\bm{\beta}$ coincides with $\bm{\pi}^c$ (with an interchange of axes $1\lra2$ and charge inversion):
\beq
\beta_i=-\frac{l_B^2}{\hbar}\eps^{ij}\pi_j^c\,,\quad\mbox{or}\quad
(\beta_1,\beta_2)=(-\frac{l_B^2}{\hbar}\pi_2^c,\frac{l_B^2}{\hbar}\pi_1^c)\,.  \label{beta2pi}
\eeq
Consequently, the representation of the MT operator \eqref{symTR} becomes
\beq
\Tm{R}=\Exp{\frac{i}{\hbar}\bm{R}\cdot\bm{\pi}^c} =\Exp{\frac{i}{l_B^2}(R_1\beta_2-R_2\beta_1)} \,, \label{symTRbeta}
\eeq
showing that, as in the Landau gauge (see \eqref{LandTR}), $\bm{\beta}$ serves as the generator of MT group.

%\vskip\baselineskip
%%%%%%%%%%%%%%%%%%%%%%%%%%%%%%
%   A.2   gauge reverting 
%%%%%%%%%%%%%%%%%%%%%%%%%%%%%%
\subsection{Reverting from the symmetric gauge}
\indent

In the symmetric gauge, covariant derivatives may sometimes appear to be replaced by free electron derivatives, which requires careful attention. For example, the scale transformation operator $D=\bm{x}\cdot\bm{p}$ in the symmetric gauge can be written as:
\beq
D:=-i\hbar(x\partial_1+y\partial_2)=\bm{x}\cdot\bm{\pi}=\bm{x}\cdot\bm{\pi}^c\,.
%=i\hbar(a^\dagger b^\dagger-ab+1)  \label{Dinv0}
\eeq
Here, the gauge-invariant covariant derivative part is replaced by that of a free electron, making $D=\bm{x}\cdot\bm{p}$ seemingly inconsistent. This relationship indicates that $\bm{x}\cdot\bm{\pi}$ represents an extension to the gauge-invariant form $\mathcal{D}$ of $D$. In other words, choosing the symmetric gauge suggests the existence of a gauge-invariant operator for which $\mathcal{D}=D$.

The operator $D$ satisfies the commutation relation $[D,p_i]=i\hbar p_i$, which corresponds to the first relation in the conformal algebra \eqref{DPKalg}, in the absence of gauge potential or for neutral charge. In the symmetric gauge, however, using \eqref{beta2pi}, one can show that
\beq
[D,\pi^c_j]=i\hbar\pi_j\,,\quad
[D,\pi_j]=i\hbar\pi_j^c=i\frac{\hbar^2}{l_B^2}\eps^{jk}\beta_k\,.  \label{Dpi0}
\eeq

The rotation operator requires similar considerations. Generally, the angular momentum operator
\begin{align}
&\bm{J}=\bm{x}\times\bm{p}\,\quad\mbox{or}\quad J_i=-i\hbar l_i\,,\quad\mbox{with}\quad
l_i=\sum_{j,k}\eps^{ijk}x_j\partial_k   \label{defAM}\\
&J_\pm=J_1\pm i J_2\,,\quad \bm{J}^2=J_3^2+\frac{1}{2}(J_+J_-+J_-J+)
\end{align}
satisfies:
\begin{align}
& [J_i,J_j]=i\hbar\eps^{ijk}J_k\,,\quad  [J_i,\bm{J}^2]=0   \label{JJalg1}  \\
&[J_+,J_-]=2\hbar J_3\,,\quad   [J_3,J_\pm]=\pm\hbar J_\pm \,. \label{JJalg2}
\end{align}
In the present system, since the $z$-coordinate is absent, the angular momentum components $J_1$ and $J_2$ do not exist, and only $J_3$ can be defined. Thus, the associated commutation relations become trivial. When the symmetric gauge is employed, 
$J_3$ takes the form
\beq
J_3:=-i\hbar(x\partial_2-y\partial_1)=\hbar(a^\dagger a-b^\dagger b)     \label{J3inv}
\eeq
and its commutation relations with $\bm{\pi}$ are given as follows. 
Consequently, $J_3$ commutes with the Hamiltonian $H$:
\beq
[J_3,\pi_j]=i\hbar\eps^{3jk}\pi_k\,,\quad [J_3,H]=0\,.  \label{J3pi}
\eeq
The first equation corresponds to the Poincar\'{e} subalgebra \eqref{MXalg} within the conformal group. Although $J_3$ itself is not gauge-invariant, the expression \eqref{J3inv} represents a gauge-invariant harmonic oscillator form (since it can be written in terms of $\pi_i$). Following the same reasoning as for $\mathcal{D}$, this suggests the existence of a gauge-invariant counterpart $\mathcal{J}_3$ associated with $J_3$, such that $\mathcal{J}_3=J_3$ in the symmetric gauge. Moreover, $\mathcal{J}_3$ is expected to satisfy \eqref{J3pi}.

\medskip
When choosing a gauge other than the symmetric gauge, additional terms appear in the right-hand side of \eqref{J3inv} (after the second equality). However, these terms cancel with phase factors from the wave function during gauge transformations (see \eqref{J3Lam}). When considering gauge-invariant quantities, it is sufficient to work with the harmonic oscillator representation of $J_3$ (see Appendix~\ref{sec:gauge} for details). This means that after finding the harmonic oscillator representation in the symmetric gauge, returning to a representation in terms of $\pi$ yields a gauge-invariant form valid in any gauge. Similarly, the gauge-invariant form of the scale transformation $D$ can be found through its harmonic oscillator form.

To summarize, by determining the harmonic oscillator representation of operators in the symmetric gauge, we can revert to gauge-independent expressions using these oscillators, thereby obtaining gauge-invariant operator forms. Following this method, we define $\mathcal{D}$ and $\mathcal{J}_3$ as:
\begin{align}
&\mathcal{D}:=i\hbar(a^\dagger b^\dagger-ab+1)=\bm{x}\cdot\bm{\pi}\,,  \label{Dcal}\\
&\mathcal{J}_3:=\hbar(a^\dagger a -b^\dagger b)=(\bm{x}\times\bm{\pi})_3-\frac{\hbar}{2l_B^2}\bm{x}^2\,. \label{J3cal}
\end{align}
These satisfy the same relations as \eqref{Dpi0} and \eqref{J3pi}, which can be verified:
\begin{align}
&[\mathcal{D},\pi_j]=i\frac{\hbar^2}{l_B^2}\eps^{jk}\beta_k\,,  \label{calDpi}  \\
&[\mathcal{J}_3,\pi_j]=i\hbar\eps^{3jk}\pi_k\,,\quad [\mathcal{J}_3,H]=0\,. \label{calJ3pi} 
\end{align}
Their commutation relations with $\bm{\beta}$ are
\beq
[\mathcal{D},\beta_j]=-il_B^2\eps^{jk}\pi_k \,,\quad
[\mathcal{J}_3,\beta_j]=i\hbar\eps^{jk}\beta_k \,.   \label{DJbeta}
\eeq
Of course, in the symmetric gauge, $\mathcal{D}=D$ and $\mathcal{J}_3=J_3$. $\mathcal{J}_3$ can also be written as:
\beq
\mathcal{J}_3=\frac{1}{2}(\tilde{\bm{x}}\times\bm{\pi})_3-\frac{\hbar}{2l_B^2}\bm{\beta}^2\,,\quad 
\tilde{\bm{x}}=\bm{x}-\bm{\beta}\,.
\eeq

$\mathcal{J}_3$ and $\mathcal{D}$ can also be expressed in forms that do not explicitly involove the coordinate $x$:
\beq
\mathcal{D}=\bm{\beta}\cdot\bm{\pi}+i\hbar\,,
\eeq
\begin{align}
\mathcal{J}_3&=\frac{1}{\omega}H-\frac{\hbar}{2l_B^2}\bm{\beta}^2=\frac{\hbar}{2}(\frac{l_B^2}{\hbar^2}\bm{\pi}^2
-\frac{\hbar^2}{l_B^2}\bm{\beta}^2)     \label{HisJ3beta}  \\
&=\frac{\hbar}{2}(\frac{l_B}{\hbar}\bm{\pi}+\frac{\hbar}{l_B}\bm{\beta})(\frac{l_B}{\hbar}\bm{\pi}-\frac{\hbar}{l_B}\bm{\beta}) \,.
\end{align}

%%%%%%%%%%%%%%%%%%%%%%%%%%%%%%%%%%%%%%%%%%%%%%%%%%%%%%%%%%%%%%%%%%%%
%                        B.   Gauge invariance of J_3
%%%%%%%%%%%%%%%%%%%%%%%%%%%%%%%%%%%%%%%%%%%%%%%%%%%%%%%%%%%%%%%%%%%%
\setcounter{equation}{0}
\section{Gauge reverting invariance of $J_3$ and $D$}\label{sec:gauge}
\indent

To formally understand how the method in Appendix~\ref{sec:Landau} is equivalent to performing gauge transformations, we verify gauge invariance by explicitly applying gauge transformations in this appendix.

We begin by reviewing the gauge invariance of the Hamiltonian (i.e., confirming that $H\psi = E\psi$ is invariant). The gauge transformations of the gauge potential, wavefunction, and Hamiltonian are given by:
\begin{align}
&A_i\ra A'_i=A_i-\frac{\hbar c}{e}\partial_i\Lambda \\
&\psi\ra\psi'=e^{i\Lambda}\psi \\
&H\ra H'=\frac{1}{2m}\sum_i (\pi_i-\hbar\partial_i\Lambda)^2\,.
\end{align}
Since no scalar potential is present, we assume $\partial_t\Lambda = 0$ so that the wave equation $i\hbar\partial_t\psi=H\psi$ remains invariant. Considering $H'\psi'$, we first apply the first-order differential operator to the wavefunction:
\[
(\pi_i-\hbar\partial_i\Lambda)\psi'=(p_i+\frac{e}{c}A_i-\hbar\partial_i\Lambda)e^{i\Lambda}\psi=e^{i\Lambda}(p_i+\frac{e}{c}A_i)\psi\,.
\]
Applying $\pi_i-\hbar\partial_i\Lambda$ once more leads to the transformed Hamiltonian:
\beq
H'\psi'=e^{i\Lambda}H\psi\,, \label{Hpsi}
\eeq
indicating that the effect of the gauge transformation appears only as a phase factor in the eigenfunction. This ensures that the eigenvalue remains gauge-invariant:
\beq
<\psi'|H'|\psi'>=<\psi|e^{-i\Lambda}H'|\psi'>=<\psi|H|\psi>\,. \label{Hexp}
\eeq

Next, we examine the reverting transformation of angular momentum $J_3$ from the symmetric gauge $A^{sym}$. Since $\mathcal{J}_3 = J_3$ in the symmetric gauge, explicitly indicating gauge dependence as
\beq
J_3=\mathcal{J}_3[A_i^{sym}]\,,\quad J'_3=\mathcal{J}_3[A'_i]\,, \label{Jsym} 
\eeq
we have the transformation from $J_3$ using \eqref{J3cal} as follows:
\begin{align}
A^{sym}_i\ra A'_i&=-\frac{1}{2}B\eps^{ij}x_j-\frac{\hbar c}{e}\partial_i\Lambda \\
\psi\ra\psi'&=e^{i\Lambda}\psi   \label{psiLam}   \\
J_3\ra J'_3 &=\hbar(a^\dagger a-b^\dagger b)' \nn\\
&=J_3-\frac{\hbar}{2l_B^2}(x^2+y^2)+\frac{\hbar}{B l_B^2} (xA'_2-yA'_1)  \nn\\
&=J_3-\hbar(x\partial_2\Lambda-y\partial_1\Lambda)\,.  \label{J3Lam}
\end{align}
The second term in $J'_3$ cancels the phase shift in the eigenfunction, ensuring eigenvalue invariance. From \eqref{psiLam} and \eqref{J3Lam}, we obtain
\beq
J'_3\psi'=J_3e^{i\Lambda}\psi-\hbar(x\partial_2\Lambda-y\partial_1\Lambda)e^{i\Lambda}\psi\,, \label{J3psi}
\eeq
where the second term on the RHS cancels with $[J_3,e^{i\Lambda}]\psi$. In fact:
\begin{align}
J_3e^{i\Lambda}\psi&=-i\hbar(x\partial_2-y\partial_1)e^{i\Lambda}\psi \nn\\
&=\hbar(x\partial_2\Lambda-y\partial_1\Lambda)e^{i\Lambda}\psi+e^{i\Lambda}J_3\psi\,.
\end{align}
Substituting this into the first term of \eqref{J3psi} yields:
\beq
J'_3\psi'=e^{i\Lambda}J_3\psi\,,
\eeq
showing that the transformation effect appears only as a phase in the eigenfunction in the same form as \eqref{Hpsi}. Therefore, similarly to \eqref{Hexp}, the eigenvalue is invariant:
\beq
<\psi'|J'_3|\psi'>=<\psi|e^{-i\Lambda}J'_3|\psi'>=<\psi|J_3|\psi>\,.
\eeq

In general, an operator $\hat{\mathcal{O}}$ is gauge-invariant if it satisfies either of the following:
\begin{align}
\hat{\mathcal{O}}'\psi'&=e^{i\Lambda}\hat{\mathcal{O}}\psi \\
\hat{\mathcal{O}}\ra \hat{\mathcal{O}}'&=e^{i\Lambda}\hat{\mathcal{O}}e^{-i\Lambda}
\label{Otrans2}
\end{align}
or
\beq
\hat{\mathcal{O}}(\hat{\pi}_i)  \ra \hat{\mathcal{O}}'(\hat{\pi}_i)=\hat{\mathcal{O}}(\hat{\pi}_i)+
\hat{\mathcal{O}}(\hat{\pi}_i\ra-i\hbar\partial_i\Lambda)-\hat{\mathcal{O}}(0)\,.
\label{Otrans3}
\eeq
According to \eqref{Otrans2}, any polynomial in $\hat{\mathcal{O}}$ is also gauge-invariant.

The form \eqref{Otrans3}, as seen in \eqref{J3psi}, holds when $\hat{\mathcal{O}}$ is a linear function of $\pi_i$, and follows from applying the Leibniz rule to the local phase factor $e^{i\Lambda}$. Let us show concrete examples of the reverting transformations from the symmetric gauge for $\pi_i$, $J_3$, and $D$:
\begin{align}
&\pi'_i=\pi-i\hbar\partial\Lambda\,,   \label{piLam}\\
&J'_3=J_3-\hbar(x\partial_2-y\partial_1)\Lambda\,, \\
&D'=D-\hbar(x\partial_1+y\partial_2)\Lambda\,,
\end{align}
where $D$ and $D'$ are defined as follows in the same convention as in \eqref{Jsym}:
\beq
D=\mathcal{D}[A_i^{sym}]\,,\quad D'=\mathcal{D}[A_i']=i\hbar(a^\dagger a-ab+1)'\,.
\eeq

From \eqref{piLam},  the operators $a^\dagger$, $a$, $b^\dagger$, $b$, and $\beta_i$, 
which are polynomials of $\pi_i$,
are gauge-invariant, and hence quantities such as $H$, $\mathcal{J}_3$, and $\mathcal{D}$, which are polynomials constructed from them, are also gauge-invariant.

%\newpage
\newcommand{\NP}[1]{{\it Nucl.{}~Phys.} {\bf #1}}
\newcommand{\PL}[1]{{\it Phys.{}~Lett.} {\bf #1}}
\newcommand{\Prep}[1]{{\it Phys.{}~Rep.} {\bf #1}}
\newcommand{\PR}[1]{{\it Phys.{}~Rev.} {\bf #1}}
\newcommand{\PRL}[1]{{\it Phys.{}~Rev.{}~Lett.} {\bf #1}}
\newcommand{\PTP}[1]{{\it Prog.{}~Theor.{}~Phys.} {\bf #1}}
\newcommand{\PTPS}[1]{{\it Prog.{}~Theor.{}~Phys.{}~Suppl.} {\bf #1}}
\newcommand{\MPL}[1]{{\it Mod.{}~Phys.{}~Lett.} {\bf #1}}
\newcommand{\IJMP}[1]{{\it Int.{}~Jour.{}~Mod.{}~Phys.} {\bf #1}}
\newcommand{\IJTP}[1]{{\it Int.{}~J.{}~Theor.{}~Phys.} {\bf #1}}
\newcommand{\JPA}[1]{{\it J.{}~Phys.} {\bf A}:\ Math.~Gen. {\bf #1}~}
\newcommand{\JHEP}[1]{{\it J.{}~High Energy{}~Phys.} {\bf #1}}
\newcommand{\JMP}[1]{{\it J.{}~Math.{}~Phys.} {\bf #1} }
\newcommand{\CMP}[1]{{\it Commun.{}~Math.{}~Phys.} {\bf #1} }
\newcommand{\LMP}[1]{{\it Lett.{}~Math.{}~Phys.} {\bf #1} }
\newcommand{\doi}[2]{\,\href{#1}{#2}\,}  %Note: _ to be escaped by {\_} in #2

%%%%%%%%%%%%%%%%%%%%%%%%%%%%%%%%%%%%%%%%%%%%%%%%%%%%%%%%%%%%%%%%%%%%%%%%
%                           REFERENCES                                 %
%  Book&Journal, DOI, arXiv
%%%%%%%%%%%%%%%%%%%%%%%%%%%%%%%%%%%%%%%%%%%%%%%%%%%%%%%%%%%%%%%%%%%%%%%%

\end{document}